\documentclass[10pt]{article}
\usepackage[english]{babel}
\usepackage[left=1in, right=1in]{geometry}

\usepackage{graphicx,subfigure} 
\usepackage{amsmath,amsfonts}
\usepackage{authblk}
\usepackage{booktabs}
\usepackage{color}

\usepackage{amssymb}
\usepackage{amsmath}
\newcommand\Ra{\mbox{\textit{Ra}}}  % Rayleigh number
\newcommand\Pran{\mbox{\textit{Pr}}} % Prandtl number, cf TeX's \Pr product
\newcommand\Lew{\mbox{\textit{Le}}}  % Lewis number
\newcommand\bnabla{\mathbf{\nabla}}
\newcommand\bcdot{\mathbf{\cdot}}
\newcommand{\beqa}{\begin{eqnarray}}
\newcommand{\eeqa}{\end{eqnarray}}
\newcommand{\beqt}{\begin{equation}}
\newcommand{\eeqt}{\end{equation}}

\begin{document}

%\begin{frontmatter}



\title{Numerical study of plume patterns in the chemotaxis--diffusion--convection coupling system}

\author[1,2]{Yannick Deleuze}
\author[2]{Chen-Yu Chiang}
\author[1,3,4]{Marc Thiriet}
\author[2,5]{Tony W.H. Sheu\thanks{\texttt{twhsheu@ntu.edu.tw}}}
\affil[1]{Sorbonne Universit\'es, UPMC Univ Paris 06, UMR 7598, Laboratoire Jacques-Louis Lions, F-75005, Paris, France}
\affil[2]{Department of Engineering Science and Ocean Engineering, National Taiwan University, No. 1, Sec. 4, Roosevelt Road, Taipei, Taiwan}
\affil[3]{CNRS, UMR 7598, Laboratoire Jacques-Louis Lions, F-75005, Paris, France}
\affil[4]{INRIA-Paris-Rocquencourt, EPC REO, Domaine de Voluceau, BP105, 78153 Le Chesnay Cedex}
\affil[5]{Center for Advanced Study in Theoretical Sciences, National Taiwan University, No. 1, Sec. 4, Roosevelt Road, Taipei, Taiwan}

% \author[label1,label2]{Yannick Deleuze}%\ead{yannick.deleuze@ljll.math.upmc.fr}
%  \author[label2]{Chen-Yu Chiang}%\ead{f01525007@ntu.edu.tw}
%   \author[label1,label3,label4]{Marc Thiriet}%\ead{marc.thiriet@upmc.fr}
%    \author[label2,label5]{Tony W.H. Sheu}\ead{twhsheu@ntu.edu.tw}
% 
% \address[label1]{Sorbonne Universit\'es, UPMC Univ Paris 06, UMR 7598, Laboratoire Jacques-Louis Lions, F-75005, Paris, France}
% \address[label2]{Department of Engineering Science and Ocean Engineering, National Taiwan University, No. 1, Sec. 4, Roosevelt Road, Taipei, Taiwan}
%  \address[label3]{CNRS, UMR 7598, Laboratoire Jacques-Louis Lions, F-75005, Paris, France}
%   \address[label4]{INRIA-Paris-Rocquencourt, EPC REO, Domaine de Voluceau, BP105, 78153 Le Chesnay Cedex}
%    \address[label5]{Center for Advanced Study in Theoretical Sciences, National Taiwan University, No. 1, Sec. 4, Roosevelt Road, Taipei, Taiwan}
%\cortext[cor1]{Corresponding author at: Department of Engineering Science and Ocean Engineering, National Taiwan University, Taipei, Taiwan. Tel.: +886 2 33665746; fax: +886 2 23929885}
\maketitle 

\begin{abstract}
A chemotaxis--diffusion--convection coupling system for describing a form of buoyant convection in which the fluid develops convection cells and plume patterns will be investigated numerically in this study. Based on the two-dimensional convective chemotaxis-fluid model proposed in the literature, we developed an upwind finite element method to investigate the pattern formation and the hydrodynamical stability of the system. The numerical simulations illustrate different predicted physical regimes in the system. In the convective regime, the predicted plumes resemble B\'enard instabilities. Our numerical results show how structured layers of bacteria are formed before bacterium rich plumes fall in the fluid. The plumes have a well defined spectrum of wavelengths and have an exponential growth rate, yet their position can only be predicted in very simple examples. In the chemotactic and diffusive regimes, the effects of chemotaxis are investigated. Our results indicate that the chemotaxis can stabilize the overall system. A time scale analysis has been performed to demonstrate that the critical taxis Rayleigh number for which instabilities set in depends on the chemotaxis head and sensitivity. In addition, the comparison of the differential systems of chemotaxis--diffusion--convection, double diffusive convection, and Rayleigh-B\'enard convection establishes a set of evidences that even if the physical mechanisms are different at the same time the dimensionless systems are strongly related to each other.
\end{abstract}

%\begin{keyword}


%\end{keyword}

%\end{frontmatter}


\begin{table}
\caption{Nomenclature description}
\label{tab:nomenclature}
\begin{center}
\begin{tabular}{lp{5cm}clp{5cm}}
     $c$               & concentration of oxygen                    & ~ & Greek symbols            & ~                                      \\
    $c_{\mathrm{air}}$ & concentration of oxygen in air             & ~ & $\beta_s$                & solute expansion coefficient           \\
    $D$                & diffusivity                                & ~ & $\beta_T$                & thermal expansion coefficient          \\
    $\mathsf{G}$       & growth rate of the plume amplitude         & ~ & $\kappa$                 & bacterial oxygen consumption rate      \\
    $h$                & container height                           & ~ & $\ell$                   & dimensionless container width    \\
    $\mathsf{H}$       & chemotaxis head                            & ~ & $\mu$                    & dynamic viscosity                      \\
    ${\mathbf j}$         & vertical unit vector upwards   & ~ & $\nu$                    & kinematic viscosity                    \\
    $\Lew$             & Lewis number                               & ~ & $\rho$                   & fluid density                          \\
    $n$                & areal number density of bacteria           & ~ & $\rho_b$                 & volumetric mass density of a bacterium \\
    ${\mathbf n}$         & unit outward normal vector                 & ~ & ~                        & ~                                      \\
    $n_0$              & initial number density of bacteria         & ~ & Superscripts             & ~                                      \\
    $\overline{n_0}$   & initial average number density of bacteria & ~ & '                        & dimensionless quantities               \\
    $p$                & pressure                                   & ~ & ~                        & ~                                      \\
    $\Pran$            & Prandtl number                             & ~ & Subscripts               & ~                                      \\
    $\Ra$              & Rayleigh number                            & ~ & $\bcdot_b$               & bacterium                              \\
    $\mathsf{S}$       & dimensionless chemotaxis sensitivity       & ~ & $\bcdot_c$               & critical                               \\
    $\mathsf{S}_{dim}$ & dimensional chemotaxis sensitivity         & ~ & $\bcdot_{\mathrm{conv}}$ & convection                             \\
    $s$                & solute concentration                       & ~ & $\bcdot_{\mathrm{diff}}$ & diffusion                              \\
    $t$                & time                                       & ~ & $\bcdot_m$               & mass                                   \\
    $\mathrm{T}$       & time scale of bacterium transport          & ~ & $\bcdot_O$               & oxygen                                 \\
    $T$                & temperature                                & ~ & $\bcdot_s$               & solute                                 \\
    $\mathbf{u}$          & velocity vector                            & ~ & $\bcdot_\tau$            & taxis                                  \\
    $u$                & velocity in x-direction                    & ~ & ~                        & ~                                      \\
    $v$                & velocity in y-direction                    & ~ & ~                        & ~                                      \\
    $V_b$              & volume of a bacterium                      & ~ & ~                        & ~                                      \\
    $\mathbf{x}=(x,y)$    & coordinate axes                            & ~ & ~                        & ~                                      \\
    \end{tabular}
\end{center}    
\end{table}

\section{Introduction}
\label{introduction}

Taxis refers to the collective motion of cells or an organism in response to an attractant gradient. The nature of the attractant stimulus can be of chemical (chemotaxis), physical (baro-, electro-, magneto-, phono-, photo-, and thermotaxis), or mechanical (hapto- and rheotaxis) origins. Chemotaxis refers to cell movement primed by a external chemical signal that can either be emitted by the same population of cells or created by an external source \cite{brenner_physical_1998}. In particular, aerotaxis is related to the movement toward a gradient of increasing oxygen concentration. 
%Bacteria such as the widely studied Bacillus Subtilis are motile and possess a flagellum for directing their motion. As bacteria are sensitive to oxygen, they swim to regions with higher concentration of dissolved oxygen. 

The phenomenon that couples chemotaxis, diffusion, and convection has been illustrated by experiments on suspensions of bacteria in a container filled with water \cite{hillesdon_bioconvection_1996,hillesdon_development_1995}. 
%In the initial well-stirred suspension, the oxygen is assumed to be uniformly distributed with a concentration being identical to the concentration of oxygen in the atmosphere. 
Oxygen diffuses in the container from the surface.  As bacteria consume oxygen, oxygen concentration falls everywhere except at the surface, hence creating a vertical concentration difference.  Bacteria move up to higher concentration of oxygen and quickly get densely packed below the surface in a relatively thin layer. 
%At a location sufficiently away from the surface, bacteria become quiescent due to the lack of oxygen. Because bacteria are denser than water, bacteria fall in the fluid due to buoyancy force and convection takes place and carries along fluid by viscous effects. Plumes also transport oxygen, thereby nourishing quiescent bacteria. 
Subsequently, Rayleigh--B\'enard-like instability appears near the surface. These dynamical instabilities exhibit complex convection patterns with plumes of bacteria falling in the fluid.

Chemotaxis--diffusion--convection is a particular case of the so-called bioconvection. Bioconvection is a more general term for suspensions of swimming microorganisms which are denser than the solvent fluid. Bioconvection and the different mechanisms of upswimming have been reviewed \cite{hill_bioconvection_2005}.

Mathematical modeling of chemotaxis has been introduced by authors in \cite{keller_model_1971} and \cite{patlak_random_1953}. Mathematical analysis mainly focused on pattern formation of microorganisms and blow-up phenomena in finite time. Key contributions on chemotactic collapse have been brought by authors in \cite{blanchet_two-dimensional_2006,herrero_singularity_1996,jager_explosions_1992,nagai_global_1998,velazquez_point_2004,perthame_cell_2007}. 
Similarly, angiogenesis is the motion of endothelial cells to form new blood vessels from preexisting vessels to locally supply oxygen. During tumor growth, tumor cells secrete a set of substances to attract endothelial cells \cite{chaplain_mathematical_2006,mantzaris_mathematical_2004}.
In angiogenesis models, chemotactic collapse was not observed \cite{corrias_chemotaxis_2003,corrias_global_2004}. 
In this study, the model equations (\ref{eqfluid1}-\ref{O2}) are used to describe chemotactic response of bacterial suspensions. The unstable agglomeration of bacterial cells at the surface leads to a descent of cell-rich plume patterns together with the formation of high-speed jets between counter-rotating vortices.

Formation and stability of plumes result from the balance between chemotaxis, diffusion, and convection of bacteria. The particular impact of each mechanism still needs to be understood. Chemotaxis is known to bring instability in system and leads to aggregation, but may also have a stabilizing effect.

The linear stability analysis of the chemotaxis-diffusion-convection system showed that a condition for linear instability depends on the taxis Rayleigh number $\Ra_{\tau}$ \cite{hillesdon_bioconvection_1996}. 
The taxis Rayleigh number is defined as the ratio of buoyancy and viscosity forces times the ratio of momentum and cell diffusivity (table~\ref{tab1:dimensionlessnumber}).
Below a critical value ($\Ra_{\tau} < \Ra_c$), then the process remains stable. From experiments, several stages were observed starting from the upward bacteria accumulation and leading to hydrodynamic formation of plumes. A weakly nonlinear stability analysis was conducted to investigate the stability of hexagon and roll patterns formed by the system of equations (\ref{eqfluid1}-\ref{O2}) \cite{metcalfe_bacterial_1998}. 
The hydrodynamic vortices formed by convection strengthen circulation of fluid and enhance the intake of oxygen into the solvent \cite{tuval_bacterial_2005}. Global existence for the chemotaxis--Stokes system for small initial bacterial population density was proved in \cite{duan_global_2010}. Then, global existence for the chemotaxis--Navier-Stokes system for a large initial bacterial population density as well as global existence of 3D weak solutions for the chemotaxis-Stokes equations were proved in \cite{liu_coupled_2011}. 
%table1--------------------------------------------------------------------------------
\begin{table*}
\caption{Representative dimensionless numbers involved in the double diffusive convection (DDC), chemotaxis--diffusion--convection (CDC), and Rayleigh-B\'enard convection (RBC). Subscripts $\bcdot_m$, $\bcdot_T$, $\bcdot_\tau$, $\bcdot_s$, $\bcdot_O$, $\bcdot_b$ stand for mass, thermal, taxis, solute, oxygen, and bacterium, respectively; $g$ is the acceleration due to gravity; $\beta_T$ and $\beta_s$ are the coefficients for thermal and solute expansion, respectively; $\nu$ is the kinematic viscosity.}
\label{tab1:dimensionlessnumber}
\begin{center}
%\resizebox{\linewidth}{!} {
\begin{tabular}{p{2cm}|c|c|c}
~ &\textbf{DDC} &\textbf{CDC}  &\textbf{RBC} \\
\hline
Rayleigh \newline number &$\begin{array}{c}
  \Ra_T=\dfrac{g\,\beta_T\,\Delta T\,\mathsf{L}^3}{\mathsf{D}_T\,\nu}\\
  \Ra_m=\dfrac{g\,\beta_s\,\Delta s\,\mathsf{L}^3}{\mathsf{D}_s\,\nu}
                 \end{array}$ 
  &$\Ra_{\tau}=\dfrac{g\,V_b\overline{n_0}(\rho_b-\rho)\mathsf{L}^3}{\mathsf{D}_b\,\mu}$ 
  &$\Ra_T=\dfrac{g\,\beta_T\,\Delta T\,\mathsf{L}^3}{\mathsf{D}_T\,\mu}$\\
\hline
Prandtl \newline number &$\Pran_T=\dfrac{\nu}{\mathsf{D}_T}$ &$\Pran_{\tau}=\dfrac{\nu}{\mathsf{D}_b}$  &$\Pran_T=\dfrac{\nu}{\mathsf{D}_T}$\\
\hline
Lewis \newline number &$\Lew_T=\dfrac{\mathsf{D}_T}{\mathsf{D}_s}$ &$\Lew_{\tau}=\dfrac{D_O}{D_b}$ &\\
\hline
Chemotaxis \newline sensitivity & ~ &$\mathsf{S}=\dfrac{\mathsf{S}_{dim}\,c_{\mathrm{air}}}{\mathsf{D}_b}$ &\\
\hline
Chemotaxis \newline head & ~ &$\mathsf{H}=\dfrac{\kappa\,\overline{n_0}\,\mathsf{L}^2}{c_{\mathrm{air}}\,D_b}$ &
\end{tabular}
%}
\end{center}
\end{table*}
%-------------------------------------------------------------------------------------

A detailed numerical study of the plume formation and merging that was related to the convergence of Rayleigh-B\'enard-type patterns was carried out in \cite{chertock_sinking_2012}. The shape and number of plumes can be controlled by initial bacterial population density. However, the sites of plumes have not been predicted. The convergence toward numerically stable stationary plumes for low and high initial density of cells is revealed in \cite{chertock_sinking_2012}. 

In this paper, a computational model based on the finite element method is proposed aiming at investigating the behavior of the two-dimensional chemotaxis--diffusion--convection system. We present numerical examples of different states of the system: (i) diffusion dominant, (ii) chemotaxis dominant, and (iii) convection dominant with the formation of descending plumes. Furthermore, we show that the chemotaxis sensitivity ($\mathsf{S}$) and the taxis Rayleigh number ($\Ra_{\tau}$) are two relevant parameters for the generation of instabilities. The effects of initial conditions and the initial cell density were also explored. Distinct initial settings lead to different solutions with plume patterns. Numerical tests show how the specified deterministic initial condition influences the overall behavior such as the number of plumes. 
In addition, the chemotaxis--diffusion--convection system is compared with the Rayleigh-B\'enard and double-diffuse convection systems. The dimensionless parameters introduced in previous papers were renamed for better readability (table~\ref{tab1:dimensionlessnumber}).

In section~\ref{mathematical model} of the present paper, the mathematical 
formulation of the chemotaxis--diffusion--convection system is presented. Section~\ref{numerics} gives the numerical method. Section~\ref{results} demonstrates the role of chemotaxis, in addition to a discussion on the simulation results and comparison with other models of buoyant convection. Finally, section~\ref{conclusion} presents some concluding remarks.

\section{Mathematical model}
\label{mathematical model}

A mathematical model for the chemotaxis--diffusion--convection is proposed in \cite{hillesdon_development_1995} and reads as follows:
\beqa
\rho \left(\frac{\partial{\mathbf{u}}}{\partial t} + \mathbf{u}\bcdot\bnabla\mathbf{u}\right) + \bnabla p - \mu\nabla^2\mathbf{u} = - n\,V_b g(\rho_b - \rho){\mathbf j}, 
\label{eqfluid1}\\
\bnabla\bcdot\mathbf{u} = 0, 
\label{eqfluid2}\\
\frac{\partial n}{\partial t} + 
  \bnabla\bcdot\left[\mathbf{u}n - D_b\bnabla n +
  \mathsf{S}_{dim}r(c)n\bnabla c\right] = 0,
\label{cells}\\
\frac{\partial c}{\partial t} + \bnabla\bcdot\left(\mathbf{u}c
  - D_{\mathrm{O}}\bnabla c\right)= - n\,\kappa r(c),
\label{O2}
\eeqa
where $\mathbf{u}=(u,v)$ denotes the velocity field of water (solvent), 
$p$ the pressure, $\rho$ and $\mu$ the water density and viscosity, 
$n$ the areal number density of bacteria (i.e., number of bacteria per 
unit area in a 2D space), 
$V_b$ and $\rho_b$ the volume and volumetric mass density of a bacterium, 
$c$ the concentration of oxygen, 
$V_b g (\rho_b - \rho){\mathbf j}$ the buoyancy force exerted by a bacterium 
on the fluid in the vertical direction (unit vector ${\mathbf j}$),
$\mathsf{S}_{dim}$ the dimensional chemotaxis sensitivity, 
$D_b$ and $D_{\mathrm{O}}$ the bacterium and oxygen diffusivities, 
$\kappa$ the bacterial oxygen consumption rate, and 
$r(c)$ the dimensionless cut-off function for oxygen concentration. The cut-off function $r(c)$ is defined by the step function \beqt r(c) = \left\{
\begin{array}{rl}
1  &  \textrm{ if } c > c^*,     \\
 0 &  \textrm{ if } c \leq c^*,
\end{array}
\right. \eeqt
 where $c^*=0.3$. 
 
Bacteria are slightly denser than water and are diluted in the solvent, so that we consider
$(\rho_b - \rho)/\rho\ll 1$ and $n\,V_b\ll 1$, respectively. The consumption of oxygen is proportional to the bacterial population density $n$. Both $n$ and $c$ are advected by the fluid. When the oxygen concentration is lower than a threshold, the bacteria become quiescent \cite{hillesdon_development_1995}, that is, they neither consume oxygen nor swim toward sites of higher oxygen concentrations. 
%The stirring motion set by the cell movement is neglected. 
The dynamics of space filling, intercellular signaling, and quorum sensing are also neglected.

%Fig00--------------------------------------------------------------------------------
\begin{figure}
\center
\includegraphics[width=0.7\linewidth]{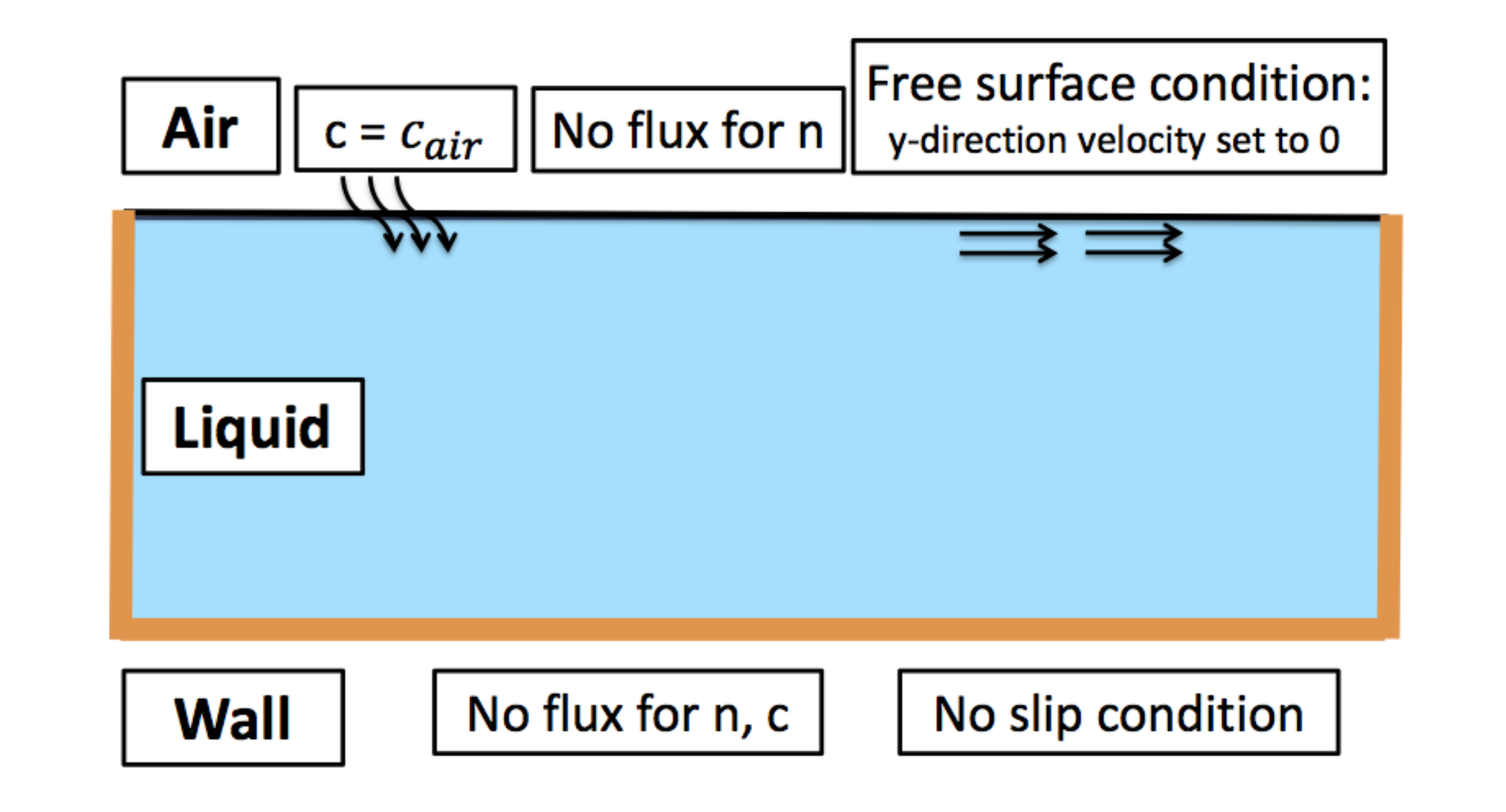}
\caption{
Boundary conditions for the system of equations (\ref{eqfluid1}-\ref{O2}). 
The air--water interface, where the oxygen concentration is equal to that 
of air, is not crossed by bacteria; the fluid vertical velocity component 
equals zero and the fluid is assumed to be free of tangential stress. 
The container walls are impermeable to bacteria and oxygen; a no-slip 
condition is imposed.}
\label{fig:BCdiag} 
\end{figure} 
%-------------------------------------------------------------------------------------

The system of equations (\ref{eqfluid1}-\ref{O2}) with the boundary conditions introduced in previous papers
(e.g. \cite{hillesdon_bioconvection_1996,tuval_bacterial_2005,chertock_sinking_2012}) is solved in a two-dimensional rectangular container ($\Omega$). The top boundary ($\Gamma_{top}$) represents the interface between liquid and air. On the free surface the concentration of oxygen is equal to the air concentration of oxygen ($c_{\mathrm{air}}$) and the free tangential stress condition as well as the absence of bacterial flux are prescribed (figure~\ref{fig:BCdiag}). Therefore:
\begin{equation}
\begin{aligned}
& \frac{\partial u}{\partial y} = 0, \; 
v = 0, \;
c = c_{\mathrm{air}}, \; \\
& \mathsf{S}_{dim}\,n\,r(c)\,\bnabla c \bcdot {\mathbf n}- D_b\,\bnabla n\bcdot {\mathbf n} = 0 \;
\textrm{ on }\Gamma_{top},
\end{aligned}
\end{equation}
where ${\mathbf n}$ is the unit outward normal vector. On the container walls ($\Gamma_{w}$), a no-slip boundary condition is prescribed and the fluxes of bacteria and oxygen equal zero:
\begin{equation}
u = 0, \;
v = 0, \;
\bnabla n \bcdot {\mathbf n}= 0, \; 
\bnabla c \bcdot {\mathbf n}= 0 \;
\textrm{ on }\Gamma_{w}.
\end{equation}
A no-slip condition at the air--water interface would enable the formation of hydrodynamic instabilities. The effect of a moving boundary due to the advection caused by an incompressible fluid flow is to be explored.

\section{Computational model}
\label{numerics}

\subsection{Scaling and setting for numerical simulations }
\label{adimensionalisation}

The characteristic length is defined by the container height $h$ and the characteristic bacterial density by the average of the initial bacterial population density defined as
\begin{align} 
\overline{n_0} &:= \frac{1}{|\Omega|}\,\int_{\Omega}\,n_0(x,t)\,dx.  
\label{n0} 
\end{align} 
This particular choice of characteristic bacterial density allows us to easily measure the total number of bacteria in each simulation for different initial distributions of bacteria. 
%A set of various spatial distributions that can lead to convective instabilities are illustrated in section~\ref{plumes}. 
 
Dimensionless variables are defined as
\beqt
\begin{aligned}
& \mathbf{x}' = \frac{\mathbf{x}}{h}, \;
t'=\frac{t}{h^2/D_b}, \;
n' = \frac{n}{\overline{n_0}}, \; \\
& c'= \frac{c}{c_{\mathrm{air}}}, \;
p' = \frac{p}{\mu D_b/h^2}, \;
 \mathbf{u}'= \frac{ \mathbf{u} }{D_b/h}.
\label{nondimen}
\end{aligned}
\eeqt
Five dimensionless parameters given below characterize the hydrodynamic and chemotaxis 
transport equations:
\beqt
\begin{aligned}
& \Pran_{\tau} {=} \frac{\nu}{D_b}, \;
  \Ra_{\tau} {=} \frac{g\,V_b\overline{n_0}(\rho_b{-}\rho)\mathsf{L}^3}{\mathsf{D}_b\,\mu}, \;
  \mathsf{S} {=} \frac{\mathsf{S}_{dim} c_{\mathrm{air}}}{D_b}, \;  \\
 & \mathsf{H} {=} \frac{\kappa\,\overline{n_0}\,\mathsf{L}^2}{c_{\mathrm{air}}\,D_b}, \;
  \Lew_{\tau} {=} \frac{D_{\mathrm{O}}}{D_b} 
\end{aligned}
\eeqt
where $\Pran_{\tau}$ is the taxis Prandtl number, 
$\Ra_{\tau}$ the taxis Rayleigh number (buoyancy-driven flow), 
$\Lew_{\tau}$ the taxis Lewis number,
$\mathsf{S}$ the dimensionless chemotaxis sensitivity, and 
$\mathsf{H}$ the chemotaxis head. 
The taxis Prandtl, Rayleigh, and Lewis numbers are analogous to the respective heat and mass Prandtl, Rayleigh, and Lewis numbers in heat and mass transfer (table~\ref{tab1:dimensionlessnumber}). The chemotaxis sensitivity ($\mathsf{S}$) and head ($\mathsf{H}$) characterize the chemotaxis system. Only $\Ra_{\tau}$ and $\mathsf{H}$ depend on the characteristic length $\mathsf{L}$ and characteristic bacterial density $\overline{n_0}$.

After removing the prime in dimensionless quantities, the set of dimensionless 
equations becomes
\beqt
\begin{aligned}
\frac{\partial \mathbf{u}}{\partial t} + \mathbf{u}\bcdot\bnabla\mathbf{u} - 
  \Pran_{\tau}\nabla^2\mathbf{u} + \Pran_{\tau}\bnabla p
  &= -\Ra_{\tau}\,\Pran_{\tau}\,n\,{\mathbf j},  \\
\bnabla\bcdot\mathbf{u} &= 0,\\
\frac{\partial n}{\partial t} + \mathbf{u}\bcdot\bnabla n -
  \nabla^2 n + \mathsf{S}\,\bnabla\bcdot(r(c) n \bnabla c) &= 0,\\
\frac{\partial c}{\partial t} + \mathbf{u}\bcdot\bnabla c -
  \Lew_{\tau}\nabla^2 c &= \mathsf{H}\,n\,r(c) . 
\end{aligned}
\label{dimensionless}
\eeqt
The system (\ref{dimensionless}) is solved in a rectangular domain 
$\Omega=[-\ell,\ell]\,\times\,[0,1]$ with the initial conditions:
\beqt
\mathbf{u}(\mathbf{x},0)=\mathbf{u}_0(\mathbf{x}), \;
n(\mathbf{x},0)=n_0(\mathbf{x}), \;
c(\mathbf{x},0)=c_0(\mathbf{x}).
\eeqt
On the top of the domain $\Omega$, the dimensionless boundary conditions are prescribed as
\beqt
 \frac{\partial u}{\partial y} = 0, \;
  v = 0, \;
   \mathsf{S} \, r(c) \, n \, \bnabla c \bcdot {\mathbf n} - D_b \, \bnabla n \bcdot {\mathbf n}=0, \;
    c=1,
\label{boundarytop} 
\eeqt
while on the other boundaries, we impose the dimensionless boundary conditions as follows
 \beqt
u = 0, \;
 v = 0, \;
  \bnabla n \bcdot {\mathbf n}=0, \;
   \bnabla c \bcdot {\mathbf n} = 0.
\label{boundarywall}
\eeqt
In the following sections, we will refer to the hydrodynamic system for the first two equations in (\ref{dimensionless}) and to the chemotaxis system for the last two equations in (\ref{dimensionless}).

\subsection{Numerical methods}
\label{sec:numericalmethods}

The governing equations in (\ref{dimensionless}) are solved using the finite element method. We adopt the biquadratic quadrilateral elements for the primitive variables $\mathbf{u}$ and the bilinear quadrilateral elements for the primitive variables $p$ so as to satisfy the LBB (Lady\v{z}henskaya \cite{ladyzenskaja_mathematical_1969} - Babu\v{s}ka \cite{babuska_error-bounds_1971} - Brezzi \cite{brezzi_existence_1974}) stability condition. There are nine nodes in one biquadratic quadrilateral element and four nodes in one bilinear quadrilateral element.
In each element, the function $\phi$ can be written as $\phi=\Sigma _{i}N^{i}\phi _i$, where $\phi _i$ are the nodal unknowns.

To avoid the convective instability while solving the convection dominated flow equations, we adopt the idea of a streamline upwind/Petrov-Galerkin (SUPG) method \cite{brooks_streamline_1982}. We implemented an inconsistent Petrov-Galerkin weighted residual scheme developed in \cite{wang_development_1996}. The resulting weak formulation reads as
\beqt
\label{eq:SUPGweak}
\def\arraystretch{2.5}
\begin{aligned}
\int _\Omega [N\frac{\partial\phi}{\partial t} + W(u\frac{\partial\phi}{\partial x} + v\frac{\partial\phi}{\partial y})
+ k( \frac{\partial N}{\partial x}\frac{\partial\phi}{\partial x}
  + \frac{\partial N}{\partial y}\frac{\partial\phi}{\partial y} ) ] d\Omega \\= \int _\Omega f d\Omega
  + k( \left. N\frac{\partial\phi}{\partial x} \right |^{x_2}_{x_1}
   + \left. N\frac{\partial\phi}{\partial y} \right |^{y_2}_{y_1} ).
\end{aligned}
\eeqt

In the context of inconsistent formulation, we have two kinds of test functions. For the non-convective terms we choose the test function to be the shape function $N$.
The test function $W$ is only applied to the convective term so as to add a numerical stabilizing term. The main purpose of employing a SUPG scheme is to add an amount of streamline artificial viscosity. The test function $W$ is therefore rewritten in two parts $W=N+B$. $B$ is called the biased part. On each node $i$, the biased part is $B^{i}_j=\tau u_{j}\frac{\partial N^{i} }{\partial x_{j} } $
where $j\in\{1,2\}$ corresponds to the Cartesian coordinates and $(u_{1},u_{2})=(u,v)$.

From the exact solution of the convection--diffusion equation in one dimension, following the derivation in \cite{wang_development_1996}, the constant $\tau$ is determined as
\beqt \tau=\frac{\delta (\gamma _1)~u~h_{1}
     + \delta (\gamma _2)~v~h_{2}}{2(u^{2}+v^{2})}. \eeqt
We define $\gamma _j=\frac{u_{j}~h_{j}}{2\, k}$ with $(h_{1},h_{2})=(\Delta x,\Delta y)$ being denoted as the grid sizes.
Finally, the derived expression for $\delta (\gamma)$ is given below
\beqt
\label{eq:SUPG_tau}
 \delta(\gamma)=
 \left\{ 
 \begin{array}{lr}
  ~~\frac{2-\cosh(\gamma) - \frac{4}{\gamma}~\tanh(\frac{\gamma}{2}) + \frac{1}{\gamma}~\sinh(\gamma)}{4~\tanh(\frac{\gamma}{2})-\sinh(\gamma)
  -\frac{6}{\gamma}~\sinh(\gamma)~\tanh(\frac{\gamma}{2})}
  , & \textrm{at end-nodes},\\
  ~~\frac{1}{2}\coth(\frac{\gamma}{2})-\frac{1}{\gamma}, & \textrm{at center-nodes}.
 \end{array}
 \right.
\eeqt

For comparison purpose in section \ref{buoyancyconvections}, the double diffusion system (\ref{eqDDC}) and the Rayleigh--B\'enard system (\ref{eqRBC}) are solved using the software Freefem++ \cite{hecht_new_2013}. The code uses a finite element method based on the weak formulation of the problem. Taylor-Hood $\mathbb{P}_2$--$\mathbb{P}_1$ elements are chosen in FreeFem++. These elements are used together with a characteristic/Galerkin formulation to stabilize the convection terms.

\subsection{Numerical validation of the coupled NS-KS equations}
\label{sec:va3}
\noindent

In this validation study, the following dimensionless differential equations accounting for the coupled Keller-Segel and incompressible viscous hydrodynamic equations are solved

\begin{figure}
\centering
\subfigure[]{\includegraphics[width=0.6\linewidth]{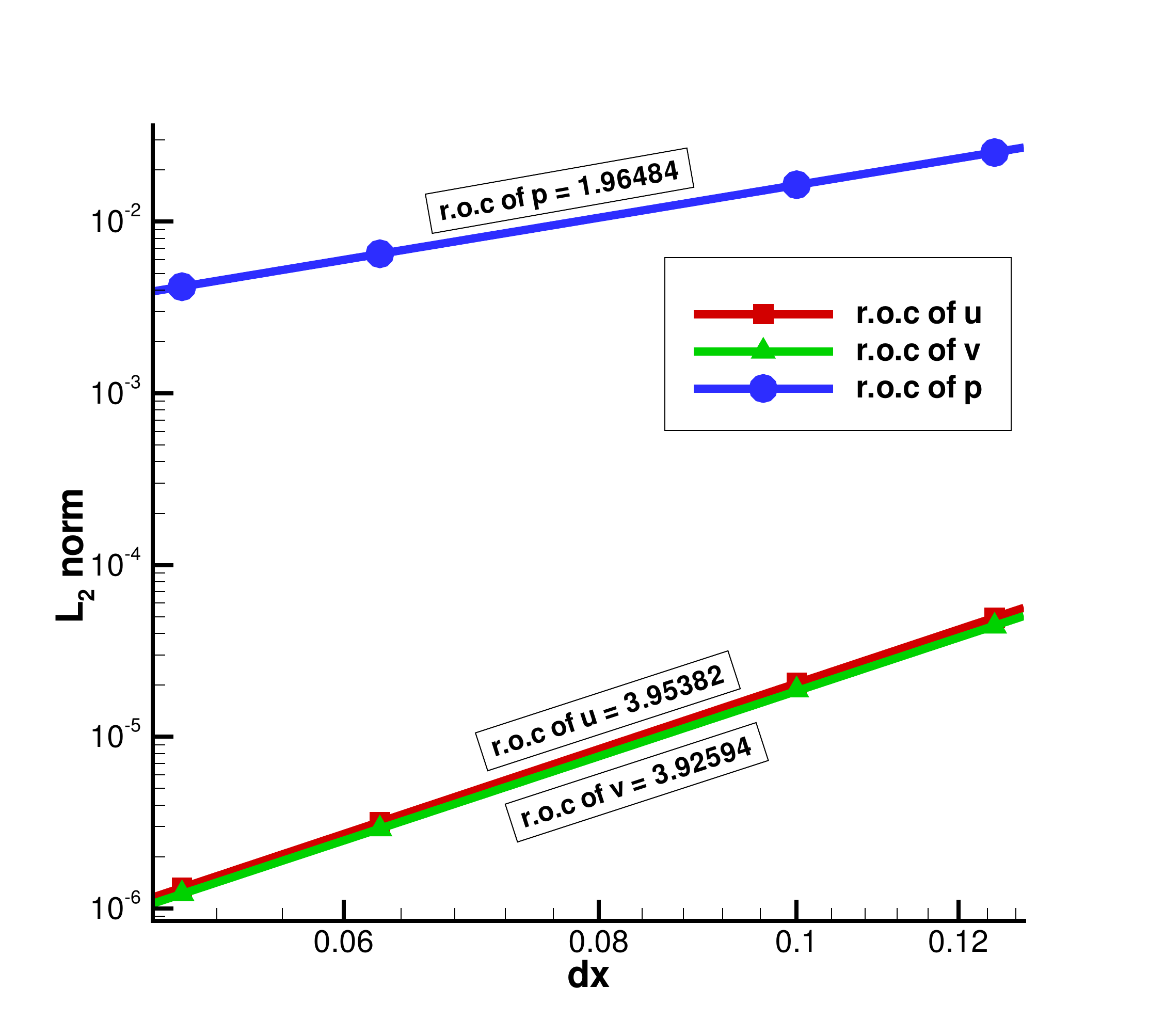}}
\subfigure[]{\includegraphics[width=0.6\linewidth]{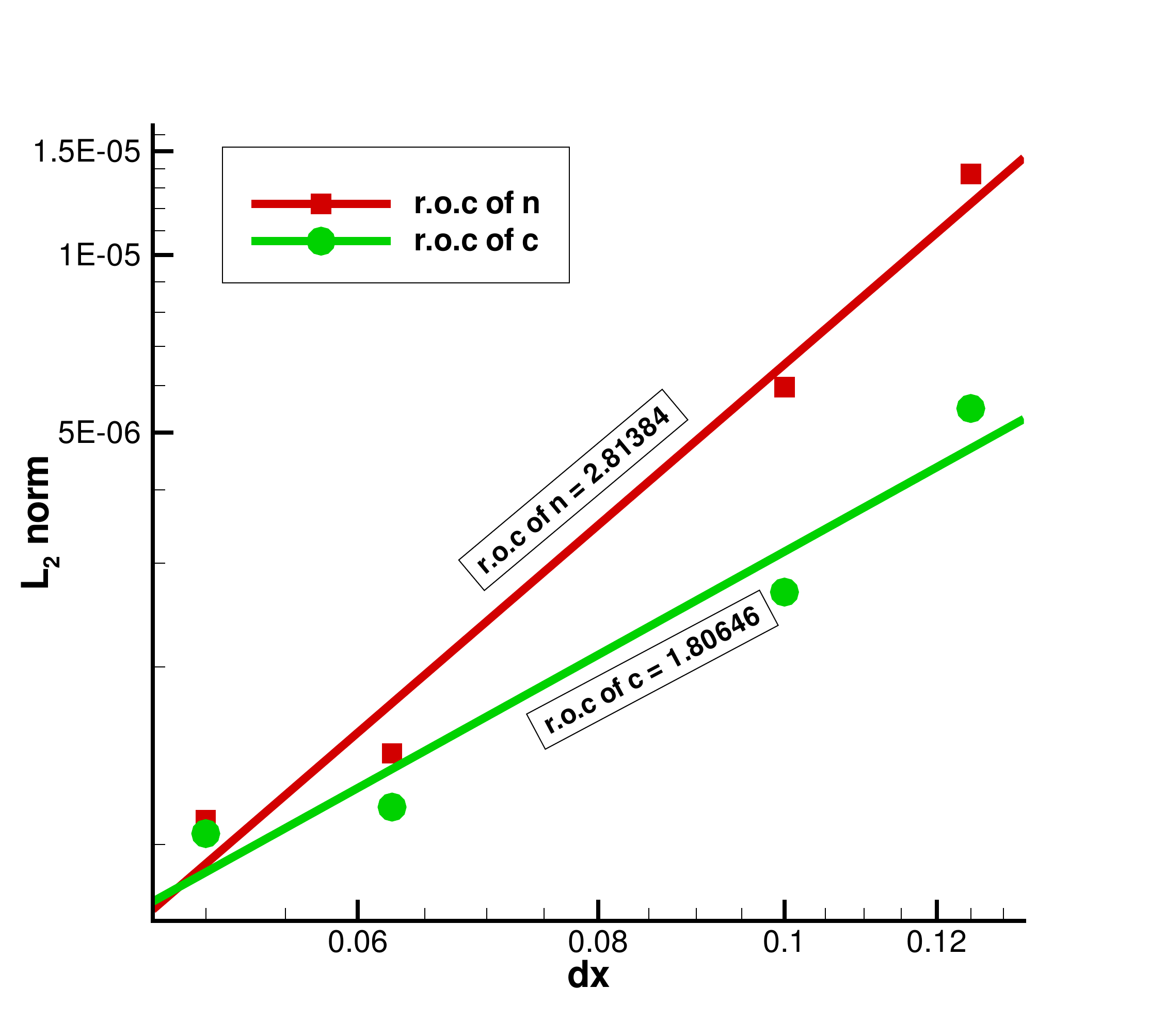}}
\caption{The computed rates of convergence (roc) for the coupled set of Navier-Stokes and Keller-Segel equations.
(a) roc= 3.95382 for $u$, roc= 3.92594 for $v$, roc= 1.96484 for $p$;
(b) roc= 2.81384 for $n$, roc= 1.80646 for $c$.}
\label{fig:ROC}
\end{figure}

\beqt
\begin{aligned}
\frac{\partial \mathbf{u}}{\partial t} + \mathbf{u}\bcdot\bnabla\mathbf{u} -
  \Pran_{\tau}\nabla^2\mathbf{u} + \Pran_{\tau}\bnabla p
  &= -\Ra_{\tau}\,\Pran_{\tau}\,n\,{\mathbf j} + \mathbf{f}_\mathbf{u},\\
\bnabla\bcdot\mathbf{u} &= 0,\\
\frac{\partial n}{\partial t} + \mathbf{u}\bcdot\bnabla n -
  \nabla^2 n + \mathsf{S}\,\bnabla\bcdot(r(c) n \bnabla c) &= f_n,\\
\frac{\partial c}{\partial t} + \mathbf{u}\bcdot\bnabla c -
  \Lew_{\tau}\nabla^2 c &= \mathsf{H}\,n\,r(c) + f_c \textrm{ in } \Omega.
\end{aligned}
\label{dimensionless}
\eeqt
The physical properties are set at the constant values of $\Pran_{\tau} = \Ra_{\tau}=\mathsf{S} = r(c) = \Lew_{\tau} = \mathsf{H} =1 $ in $\Omega = [0,1]\times[0,1]$.
Equations in (\ref{dimensionless}) are solved at $\Delta t=0.0001$ in the continuously refined four meshes with $\Delta x=\Delta y=0.125, 0.1, 0.0625, 0.05$.
The predicted errors between the simulated and exact solutions, which are $u_{exact}=-\cos(\pi x)\sin(\pi y)e^{-2\pi^2t}$, $v_{exact}=\sin(\pi x)\cos(\pi y)e^{-2\pi^2t}$, $p=c_1-0.25(\cos(2\pi x)+\cos(2\pi y))e^{-4\pi^2t} $, $n_{exact}=\cos(\pi x)\cos(\pi y)e^{-2\pi^2t}$ and $c_{exact}=\cos(\pi x)\cos(\pi y)e^{-2\pi^2t}$, are cast in their $L_2-norms$.
The source terms $\mathbf{f}_\mathbf{u}$, $f_n$, and $f_c$ are derived from the previous exact solutions.
From the predicted error norms, the spatial rates of convergence are plotted in Fig. \ref{fig:ROC}.

%Fig01--------------------------------------------------------------------------------
%-------------------------------------------------------------------------------------

The good agreement between the exact and simulated results and states of convergence demonstrate the applicability of the proposed SUPG scheme and the flow solver described in section \ref{sec:numericalmethods} to investigate the chemotactic phenomena in hydrodynamic environment.

\section{Numerical results and discussion}
\label{results}

\begin{figure}
\center
\subfigure[CDC]{\includegraphics[trim = 10mm 60mm 10mm 60mm, clip, width=\linewidth]{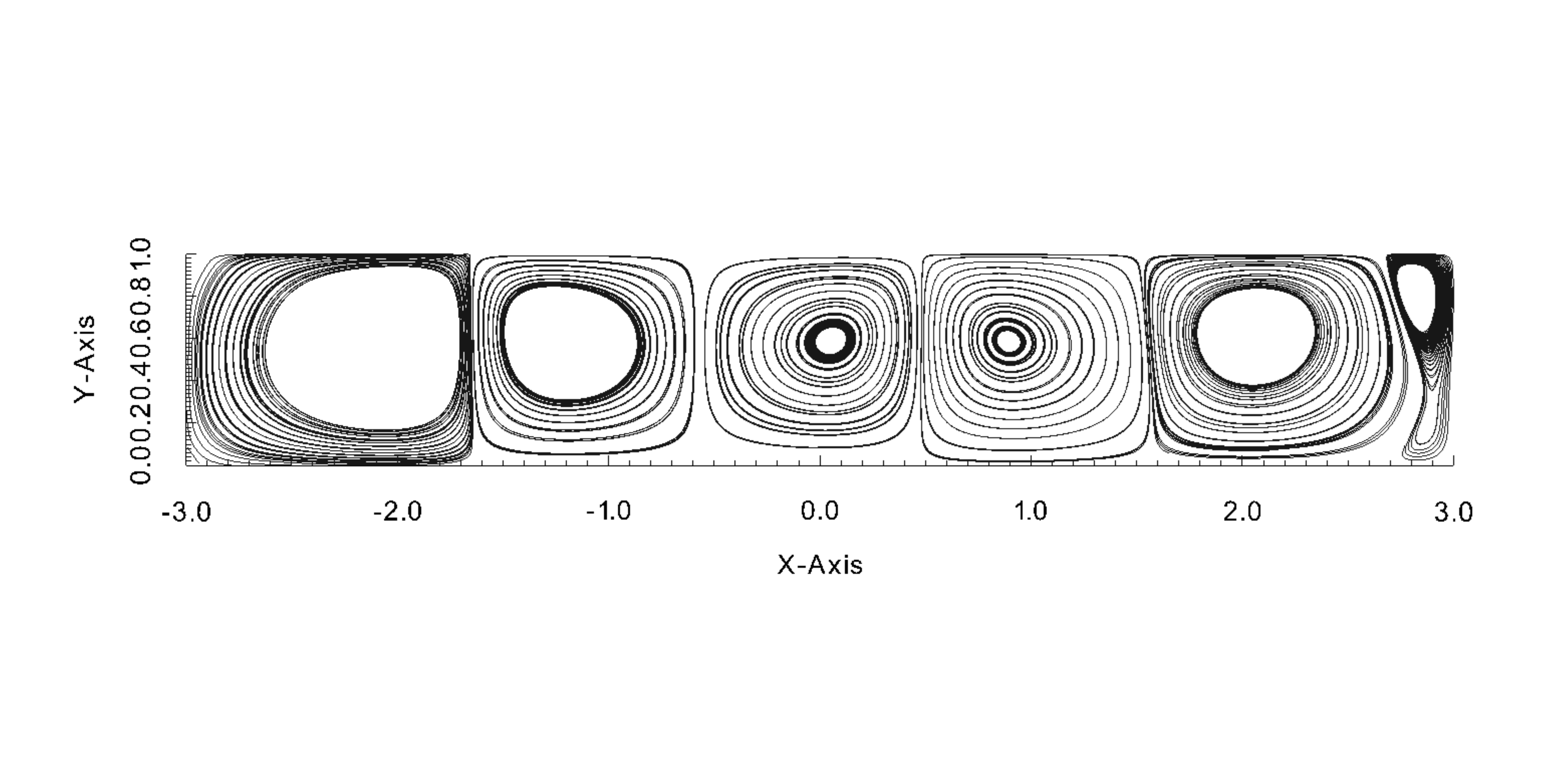}}
\subfigure[RBC]{\includegraphics[trim = 10mm 60mm 10mm 60mm, clip, width=\linewidth]{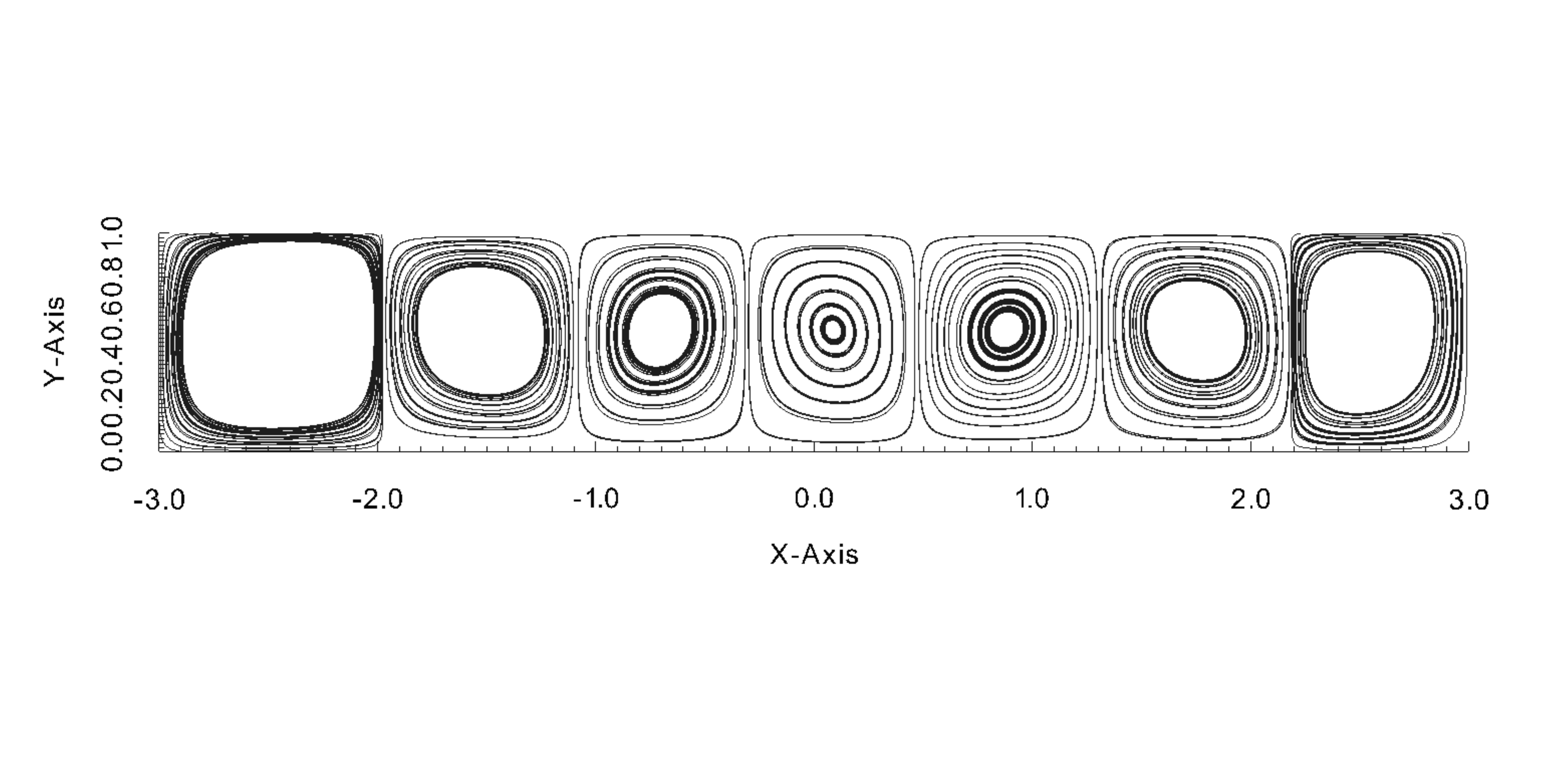}}
\subfigure[DDC]{\includegraphics[trim = 10mm 60mm 10mm 60mm, clip, width=\linewidth]{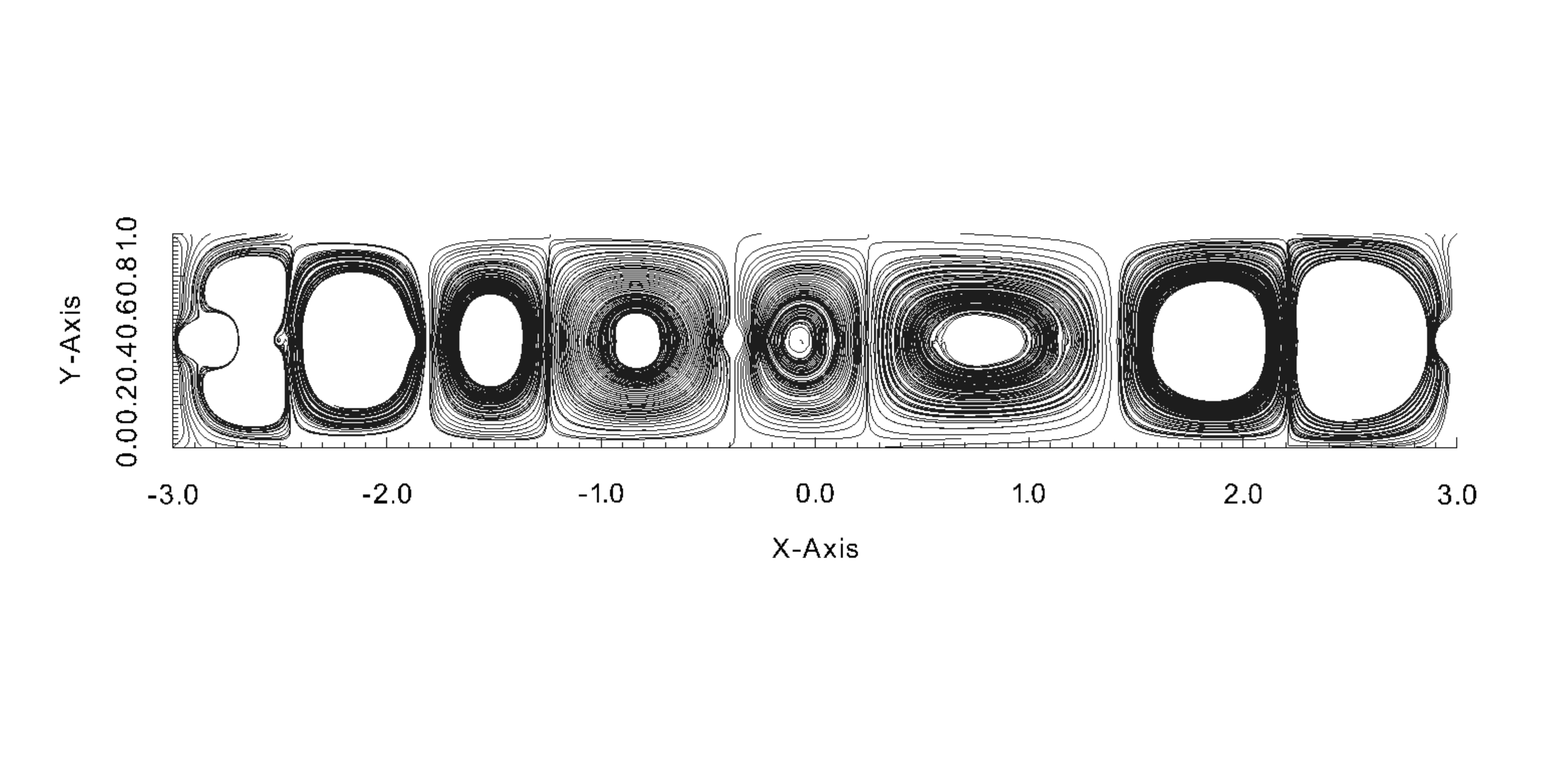}}
\caption{Examples of convection cells for the chemotaxis--diffusion--convection (CDC), Rayleigh-B\'enard convection (RBC), and double diffusive convection (DDC).}
\label{fig:convcells}
\end{figure}

The linear stability analysis of the system (\ref{dimensionless}) showed that the steady state becomes unstable for a range of physical parameters \cite{hillesdon_bioconvection_1996}. For sufficiently large characteristic bacterial density $\overline{n_0}$ and Rayleigh number $\Ra_{\tau}$, hydrodynamic instabilities appear in the region near the surface at which the bacterial density is high. This instability may be related to the Rayleigh-B\'enard instability occurring in thermal convection \cite{hillesdon_development_1995,metcalfe_bacterial_1998}. This instability develops into a descending family of bacterium-rich plumes and leads possibly to the formation of convection cells (figure~\ref{fig:convcells} (a)).

%Fig02--------------------------------------------------------------------------------
%-------------------------------------------------------------------------------------

When $\overline{n_0}$ is small, small perturbations of the velocity field are damped due to stabilizing effects of viscous friction and chemotaxis and therefore convective motion is negligible. It is shown that for low initial bacterium cell density $\overline{n_0}$ the system (\ref{dimensionless}) evolves to a steady and homogeneous state in the horizontal direction governed by the chemotaxis system \cite{chertock_sinking_2012,hillesdon_development_1995}. Like in the Keller-Segel or angiogenesis system, the random diffusion of cells in this case is balanced by the chemotaxis sensitivity of cells.

\subsection{Descending plumes}
\label{plumes}

%In the experiments presented in \cite{janosi_onset_1998}, the authors organize the experimental data into a set of seven sequential steps. The first three steps correspond to the evolving well-stirred initial state into a state of bacterial accumulation near the water-air interface. In the fourth step, the instability appears. Steps five to seven describe the evolution of the descending plumes in the convective regime.

\begin{figure*}
\center
\begin{minipage}{0.85\linewidth}
\mbox{\subfigure[t=0]{\includegraphics[width=0.48\linewidth]{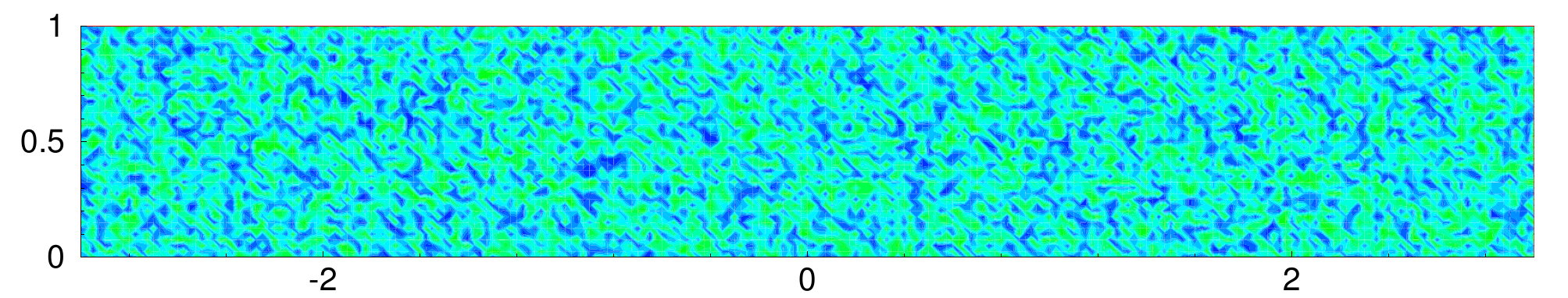}}
      \subfigure[t=0.006]{\includegraphics[width=0.48\linewidth]{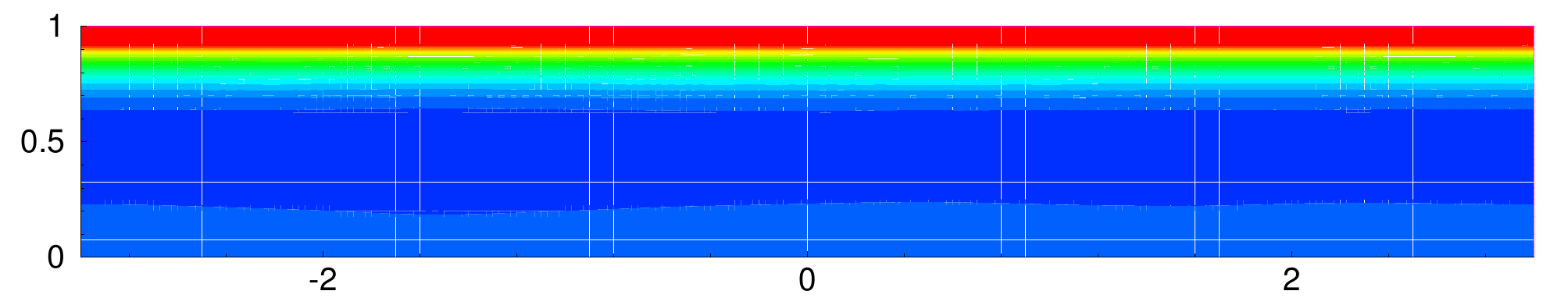}}}
\mbox{\subfigure[t=0.012]{\includegraphics[width=0.48\linewidth]{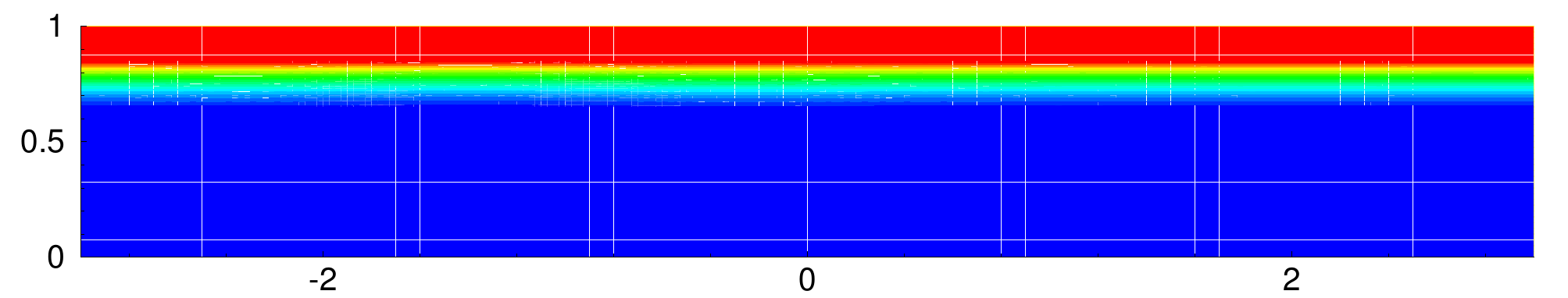}}
      \subfigure[t=0.018]{\includegraphics[width=0.48\linewidth]{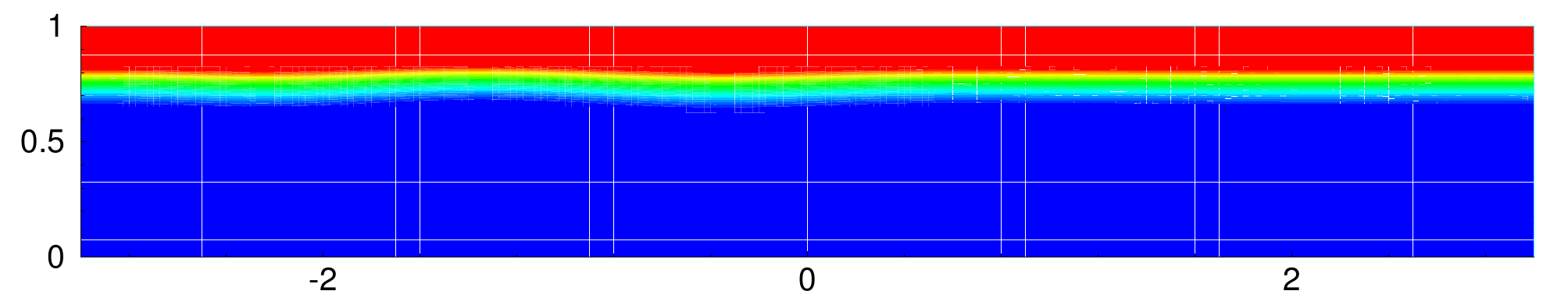}}}
\mbox{\subfigure[t=0.024]{\includegraphics[width=0.48\linewidth]{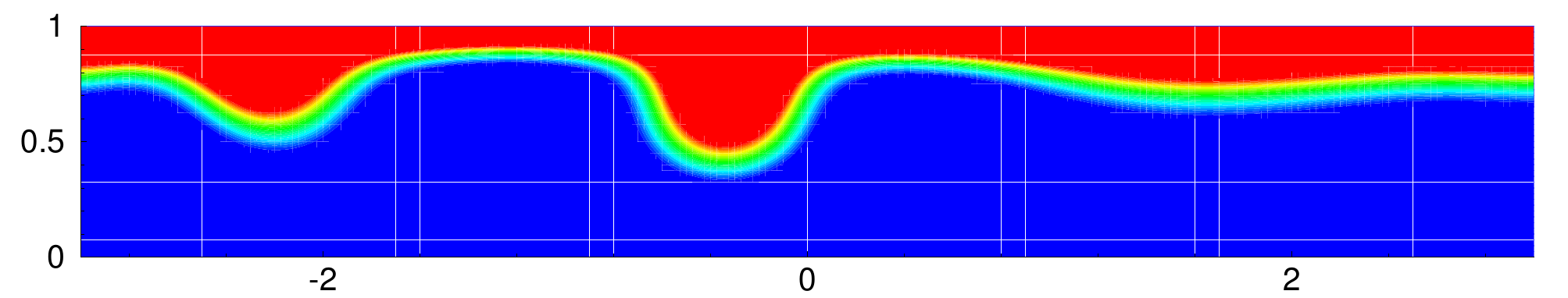}}
      \subfigure[t=0.030]{\includegraphics[width=0.48\linewidth]{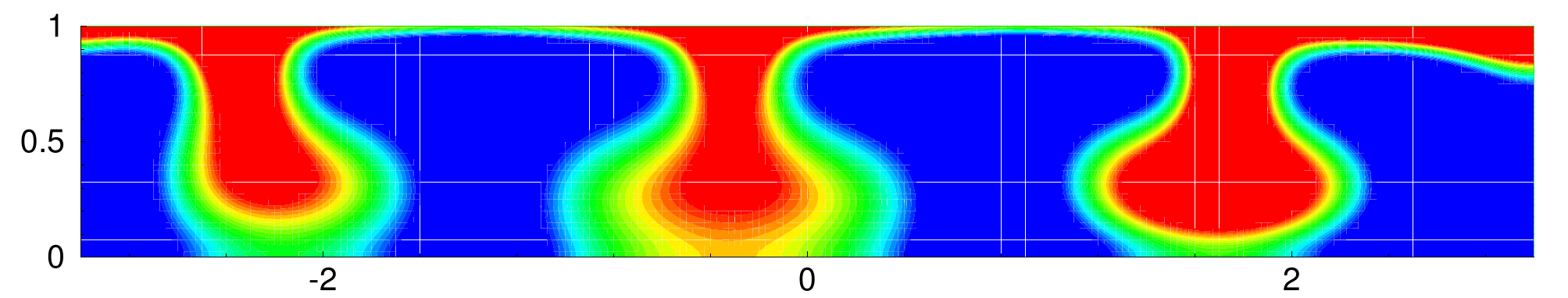}}}
\mbox{\subfigure[t=0.036]{\includegraphics[width=0.48\linewidth]{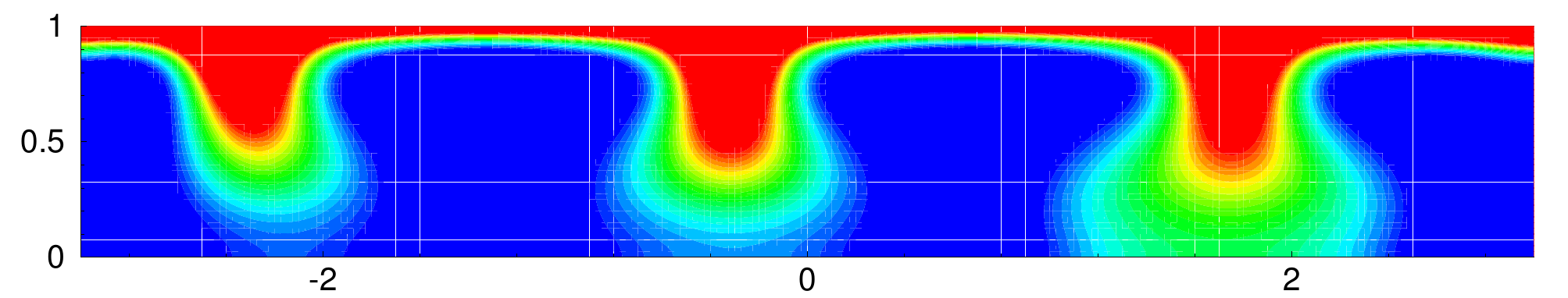}}
      \subfigure[t=0.042]{\includegraphics[width=0.48\linewidth]{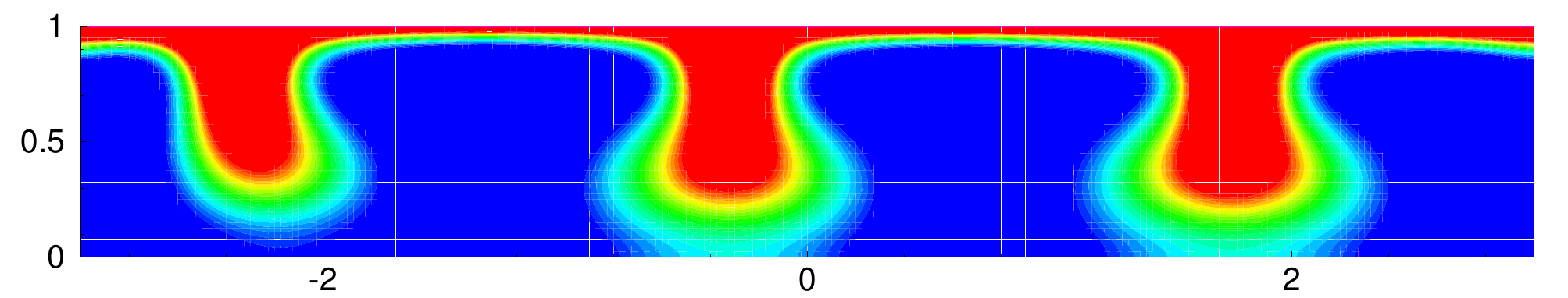}}}
\end{minipage}
\begin{minipage}{0.10\linewidth}
\mbox{\subfigure{\includegraphics[width=\linewidth]{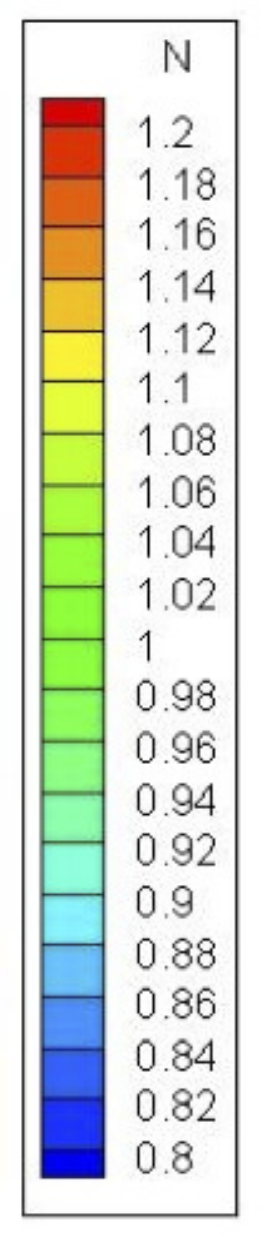}}}
\end{minipage}
\caption{Evolution of the cell density $n$ at different times. Descending plumes of bacteria develop from an initial randomly distributed bacterium population.}
\label{fig:evolution}
\end{figure*}

In the present numerical simulations, the descending plumes can be described using three phases. In the first phase (figure~\ref{fig:evolution} (a)-(c)), chemotaxis is a dominant mechanism. As bacteria consume oxygen, an oxygen concentration gradient is created that in turn provokes a bacterium motion toward the open surface where the oxygen is abundant. Bacteria chemotaxis causes the fluid to set in motion and counter rotating vortices to form (figure \ref{fig:convcells} (a)). 
Quickly, the bacterial density $n$ becomes quasi-homogeneous in the horizontal direction and is structured in layers in the vertical direction. A three-layer configuration is induced.  A layer of higher concentration of bacteria forms below the air surface: the \textit{stack layer}. When bacteria have migrated in the stack layer, a \textit{depletion layer} is generated. At the container bottom, some bacteria are inactive, because the oxygen concentration decreases below a certain threshold. An \textit{inactive layer} is established. 

%Fig03--------------------------------------------------------------------------------
%-------------------------------------------------------------------------------------

\begin{figure}
\center
	\includegraphics[width=\linewidth]{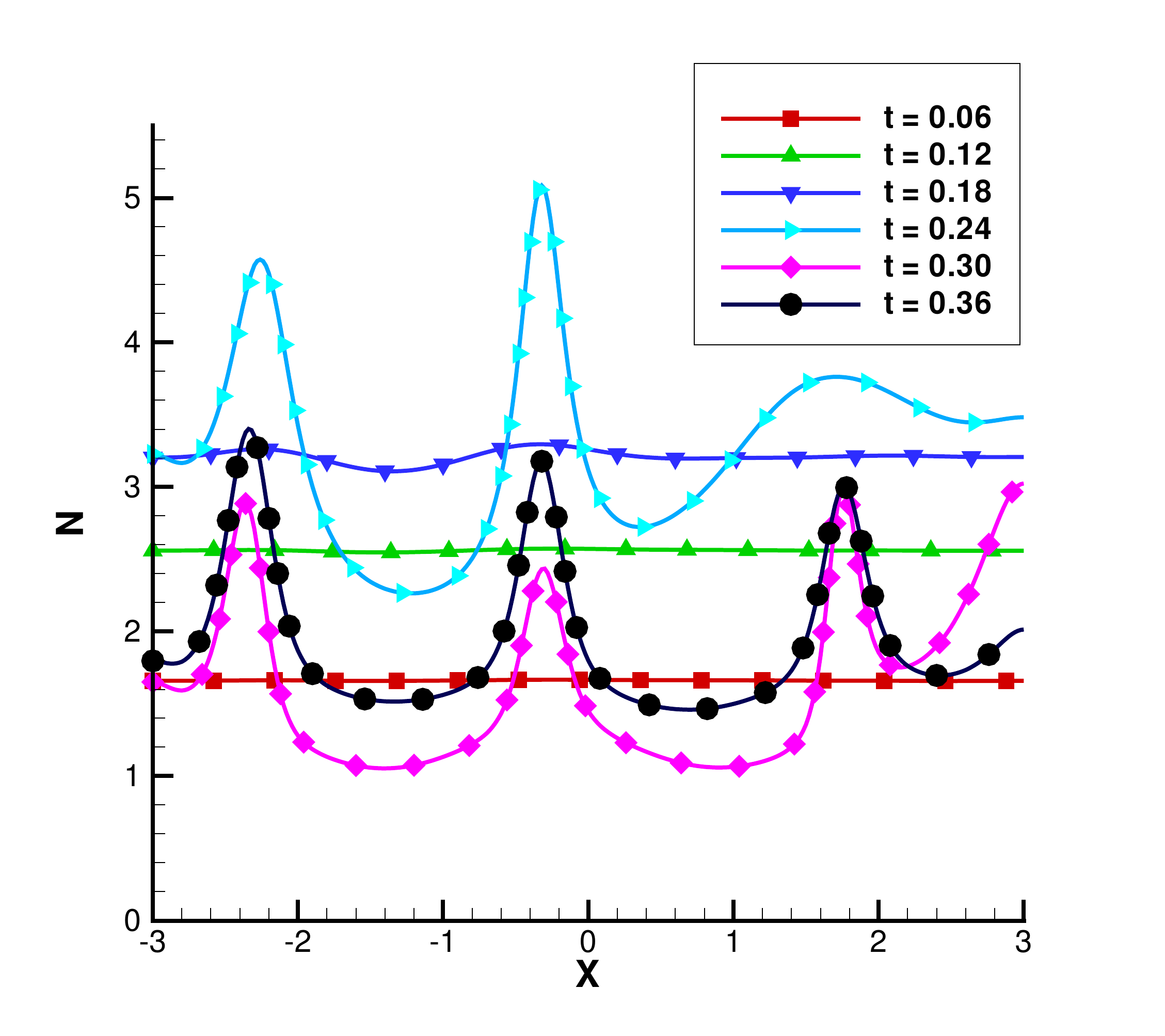}
	\caption{Evolution of the cell density number $n$ at the surface. In the initial stage of the chemotaxis--diffusion--convection, $n$ is homogeneous in the horizontal direction. As the cell density $n$ increases, hydrodynamic instability arises and the bacterial density becomes higher at some particular points.}
	\label{fig:cellsurface}
\end{figure}

The second phase (figure~\ref{fig:evolution} (d)-(e)) exhibits high bacterial density on the surface. As bacteria swim toward air-supply region, the bacterial density in the stack layer increases. Consequently, advection from the counter rotating vortices becomes significant and brings in perturbations to the stack layer (figure~\ref{fig:cellsurface}). Therefore, advection causes instabilities to occur in the bacterial density distribution in the stack layer that sets the fluid in motion.
At the same time, the oxygen concentration falls to its minimum at the bottom.

%Fig04--------------------------------------------------------------------------------
%-------------------------------------------------------------------------------------

In the third phase (figure~\ref{fig:evolution} (f)-(h)), in fluid regions with a greater bacterial density in the stack layer, buoyancy force constrains bacteria to descend in the fluid. As a result, descending plumes of bacteria develop at these particular locations. This mechanism is analogous to the Rayleigh-B\'enard instability in heat transfer problems where the fluid with a higher temperature and thus a lighter density than that above it rises to develop plumes of hot fluid.

%In summary, the time evolution of descending plumes can be divided into three main phases. In the first phase, bacteria form three layers: (1)~a superior stack layer, where bacteria agglomerate; (2)~a middle depletion layer, where bacteria migrate, attracted by oxygen; and (3)~an inferior inactive layer, where bacteria are quiescent. In the second phase, peaks of bacterial density appear at the top surface. In the final phase, perturbations at the top surface grow into the descending bacteria rich plumes.

\subsection{Stabilizing effect of chemotaxis}

When the average initial cell density $\overline{n_0}$ is large, hydrodynamic instabilities appear in the system (\ref{dimensionless}). This section is aimed at estimating $\overline{n_0}$  and dimensionless governing parameters for hydrodynamic instabilities to appear as well as at analyzing the time scales for each of the three competitive physical mechanisms: (1)~chemotaxis, (2)~diffusion, and (3)~convection of bacteria.

Convection is determined by the properties of the hydrodynamic system as well as the container height and $\overline{n_0}$ via the taxis Rayleigh number $\Ra_{\tau}$. When the Rayleigh number is below the critical value for the hydrodynamic system of interest, motion of bacteria is primarily governed by diffusion and chemotaxis; when the Rayleigh number exceeds a critical value, bacterial taxis is primarily governed by convection \cite{hillesdon_bioconvection_1996}.

Buoyancy force enables a fluid volume with a low bacterial density to ascend and a fluid volume with a high bacterial density to descend. Nevertheless, bacterium chemotaxis and friction dampen altogether the displacement.

The time scale for bacterium diffusion over the length scale $h$ is given by
\beqt
\mathrm{T}_{\mathrm{diff}} := \frac{h^2}{D_b}.
\label{timediff}
\eeqt
The buoyancy force is balanced by friction in the fluid. Therefore, the time scale for the convective displacement of bacteria over the length scale $h$ can be defined by
\beqt
\mathrm{T}_{\mathrm{conv}} := \frac{\mu}{g\,h\,\overline{n_0}\,V_b\,(\rho_b - \rho)}.
\label{timemotion}
\eeqt
The chemotaxis system is controlled by a competition between diffusion and chemotaxis of bacteria. 
%The chemotactic force is balanced by diffusion. 
Therefore, the chemotaxis time scale is expressed as
\beqt
\mathrm{T}_{\tau} := \frac{D_b}{\mathsf{S}_{dim}\,\kappa\,\overline{n_0}}.
\label{timechem}
\eeqt
In a convection-dominant process, the convection time scale is smaller than the diffusion and chemotaxis time scales (table~\ref{tab2:timescaleanalysis}). Hence:
\beqt 
\frac{g\,h^3\,\overline{n_0}\,V_b\,(\rho_b - \rho)}{D_b\,\mu}\,\equiv\,\Ra_{\tau}>1.
\label{diffconv}
\eeqt
In particular, $\Ra_{\tau}>\Ra_{c}$.  The value of the critical Rayleigh number associated with the convection $\Ra_{c}$ given by \cite{hillesdon_bioconvection_1996} in their linear stability analysis is described by the solution of an ordinary differential system. The second condition on the time scales leads to the following inequality
\beqt
\frac{g\,h\,D_b\,V_b\,(\rho_b - \rho)}{\mathsf{S}_{dim}\,\kappa\,\mu}\,\equiv\,\frac{\Ra_{\tau}}{\mathsf{S}\,\mathsf{H}}\,>\,1.
\label{chemconv}
\eeqt
Similarly to the critical Rayleigh number ($\Ra_{c}$), a critical number 
$(\mathsf{S}\mathsf{H})_{c}$ can be introduced. 

On the other hand, the chemotactic motion is predominant when the chemotaxis time scale is smaller than the diffusive time scale and convection is negligible:
\beqt
\frac{\mathsf{S}_{dim}\,\kappa\,\overline{n_0}\,h^2}{{D_b}^2}\,\equiv\,\mathsf{S}\,\mathsf{H}\,>\,1.
\label{diffchem}
\eeqt
%The chemotaxis system is controlled by a competition between diffusion and chemotaxis of bacteria. In the Keller-Segel model, collapse of the system in finite time can occur \cite{perthame_cell_2007}. In angiogenesis (equations are similar to aerotaxis (\ref{cells}-\ref{O2})), solutions of the system remain bounded \cite{corrias_global_2004}. 

%table2--------------------------------------------------------------------------------
\begin{table}
\caption{
  Phenomenological analysis based on time scales of the three competitive
  mechanisms: chemotaxis, diffusion, and convection of bacteria.  Both 
  oxygen and bacterial diffusion favor fluid homogenization.
}
\label{tab2:timescaleanalysis}
\centering
\begin{tabular}{|l|l|l|}
Dominant             &Dominant            &Dominant\\
convection           &diffusion           &aerotaxis\\
\hline
$\mathrm{T}_{\mathrm{conv}}<\mathrm{T}_{\mathrm{diff}}$
                     &$\mathrm{T}_{\mathrm{diff}}<\mathrm{T}_{\tau}$
                                          &$\mathrm{T}_{\tau}<\mathrm{T}_{\mathrm{diff}}$\\
and $\mathrm{T}_{\mathrm{conv}}<\mathrm{T}_{\tau}$;
                     &$\mathsf{S}\mathsf{H}<1$
                                          &$\mathsf{S}\mathsf{H}>1$\\
i.e., $\Ra_{\tau}>1$
                     &
                                          &\\
and $\Ra_{\tau}/\mathsf{S}\mathsf{H}>1$
                     &
                                          &\\
\end{tabular}
\end{table}
%-------------------------------------------------------------------------------------

\begin{figure*}
\begin{center}
\includegraphics[width=0.8\linewidth]{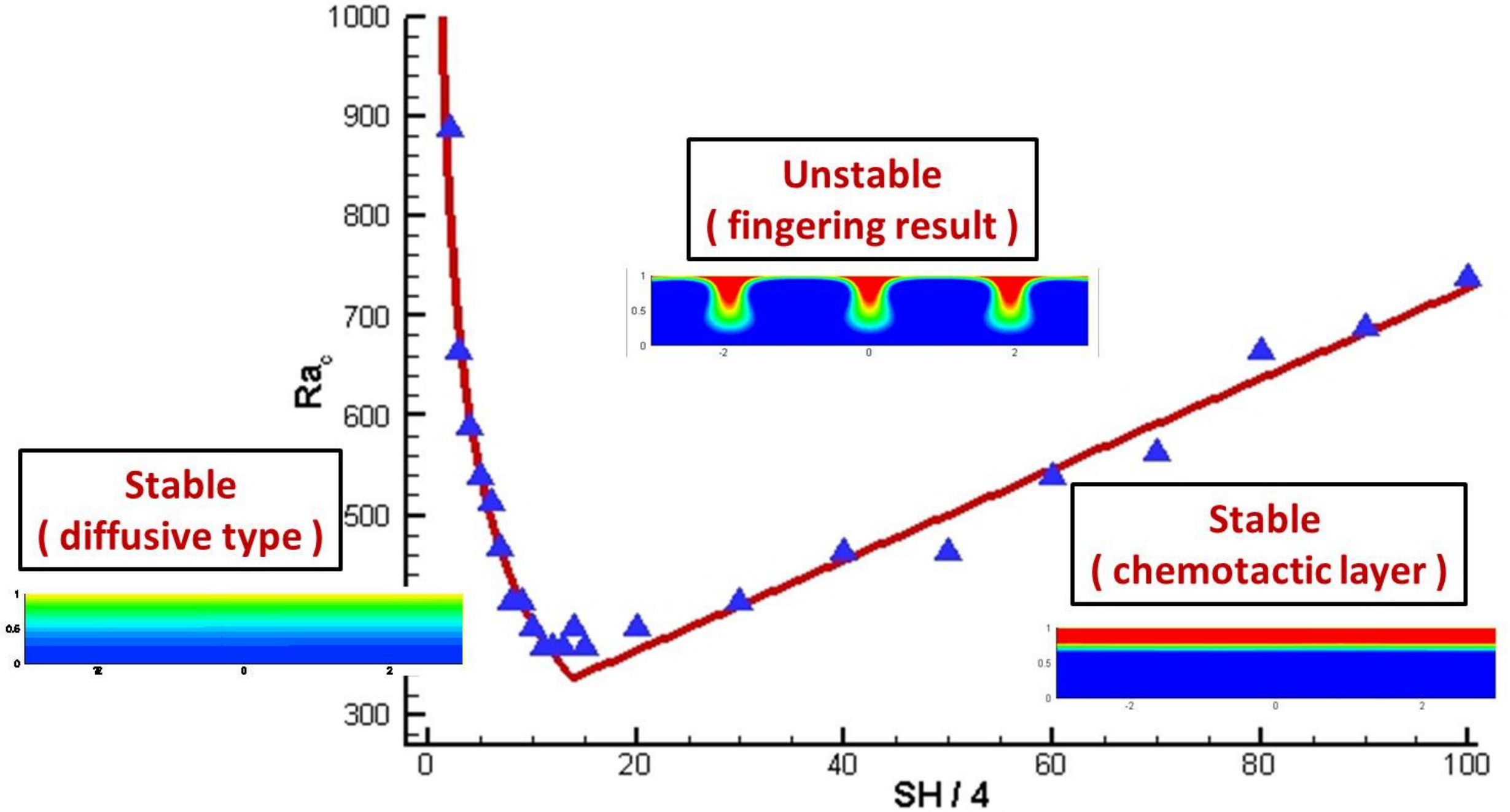}
\caption{The stable and unstable regions of the system (\ref{dimensionless}) are plotted in terms of the critical taxis Rayleigh number $\Ra_{c}$ and the product of the dimensionless chemotaxis sensitivity and chemotaxis head $\mathsf{S} \, \mathsf{H}$. The points correspond to the predicted values of $\Ra_{c}$. The red line corresponds to a fit of data in two parts: on the left side it is fitted by a power function and on the right side by a linear function. In the unstable region, the predicted solution $n$ is of the convective type (b). In the stable region the predicted solution $n$ can be of the diffusive type (a) or the chemotactic type (c).}
\label{fig:RaSP}
\end{center}
\end{figure*}

The effects of $\Ra_{c}$ and the product $\mathsf{S}\,\mathsf{H}$ were tested subject to the random initial condition. 
The results are shown in  figure~\ref{fig:RaSP} for $\Pran_{\tau} = 500$ and $\Lew_{\tau} = 5 $. 
Our numerical results agree with the linear stability analysis carried out in \cite{hillesdon_bioconvection_1996}. $\Ra_{c}$ at first falls as $\mathsf{S}\,\mathsf{H}$ increases to reach its minimum value and then rises again.  A sufficient decrease or increase of $\mathsf{S}\,\mathsf{H}$ may promote stabilization as both taxis and diffusion operate as the fluid-homogeneization factors, either directly ($\mathsf{S}\,\mathsf{H}<1$) or indirectly ($\mathsf{S}\,\mathsf{H}>1$). When $\mathsf{S}\,\mathsf{H}$ is small, stabilizing effect is due to bacterial diffusion and the solution is of the diffusive type.  When $\mathsf{S}\,\mathsf{H}$ is large, stabilizing effect results from the bacterial taxis and the solution is of the chemotactic type.

%Fig05--------------------------------------------------------------------------------
%-------------------------------------------------------------------------------------

The taxis Rayleigh number ($\Ra_{\tau}$) that characterizes the competition between diffusion and convection plays the same role as the Rayleigh number plays in classical convection. When the Rayleigh number increases, the gravitational force becomes predominant. When $\mathsf{S}\,\mathsf{H}$ rises, competition between chemotaxis and convection of bacteria becomes stronger. The condition (\ref{chemconv}) suggests that $\Ra_{c}$ increases linearly with respect to $\mathsf{S} \, \mathsf{H}$, as illustrated in figure~\ref{fig:RaSP}. We can affirm that chemotaxis has a stabilizing effect on the differential system of current interest.

%Buoyancy-driven convection is caused by a change in bacterial density in a subdomain of the fluid. As $\Ra_{\tau}$ augments, the gravitational force becomes predominant. However, its value above which instability grows and bacterial convection appears varies with the chemotaxis sensitivity of bacteria. When the product of chemotaxis sensitivity $\mathsf{S}$ and head $\mathsf{H}$ is large, $\Ra_{\tau}$ increases proportionally. We can affirm that chemotaxis has a stabilizing effect on the differential system of current interest.

\subsection{Distribution and number of plumes and initial conditions}
\label{initialcond}

\subsubsection{Position and spacing of plumes}

The exact localization of plume generation was investigated. Under the random initial condition introduced in \cite{chertock_sinking_2012}, the location is hard to be predicted. We consider deterministic initial conditions to investigate the formation and location of descending plumes. Many initial conditions among the set of tested distribution of bacterial settings are able to trigger convective patterns.

\begin{figure}
\center
\begin{minipage}{\linewidth}
\mbox{\subfigure[]{\includegraphics[width=0.45\linewidth]{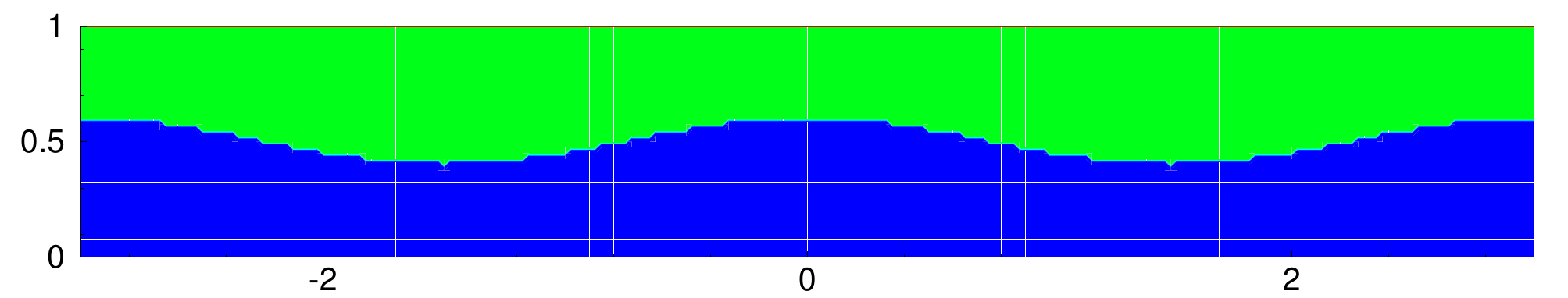}} 
\subfigure[]{\includegraphics[width=0.45\linewidth]{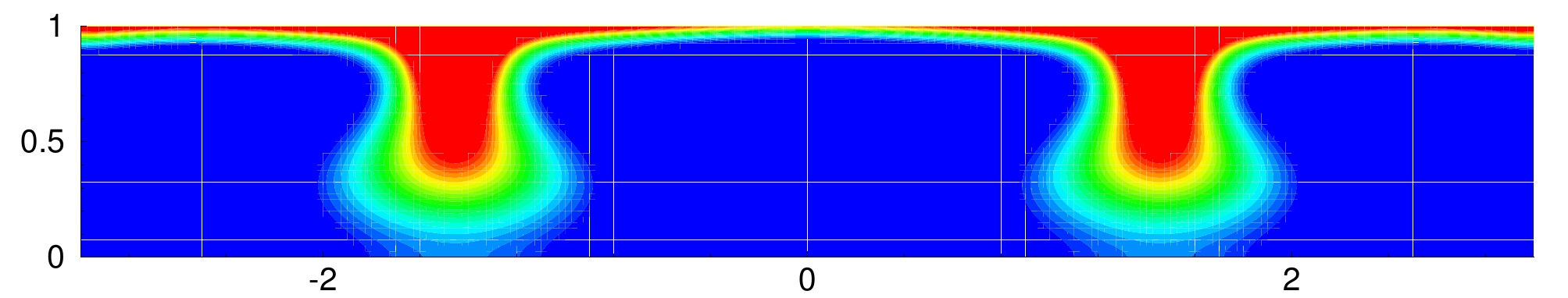}}}
\mbox{\subfigure[]{\includegraphics[width=0.45\linewidth]{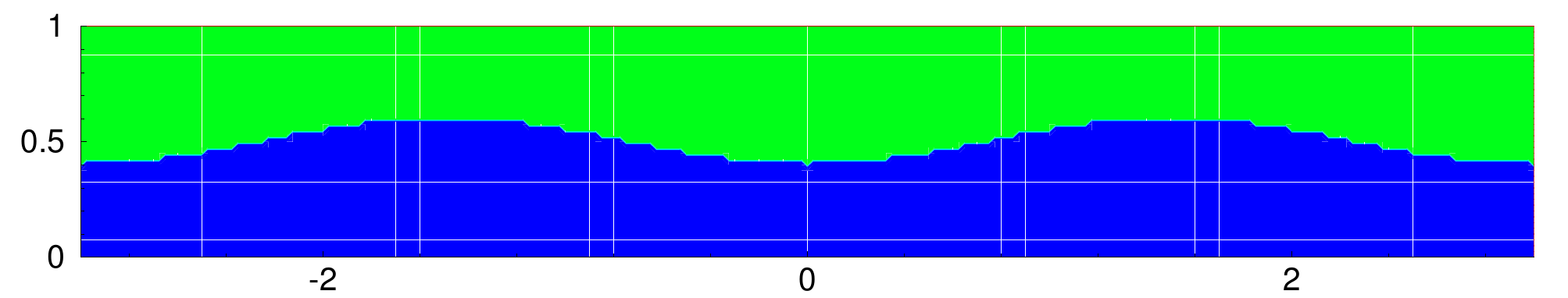}} 
\subfigure[]{\includegraphics[width=0.45\linewidth]{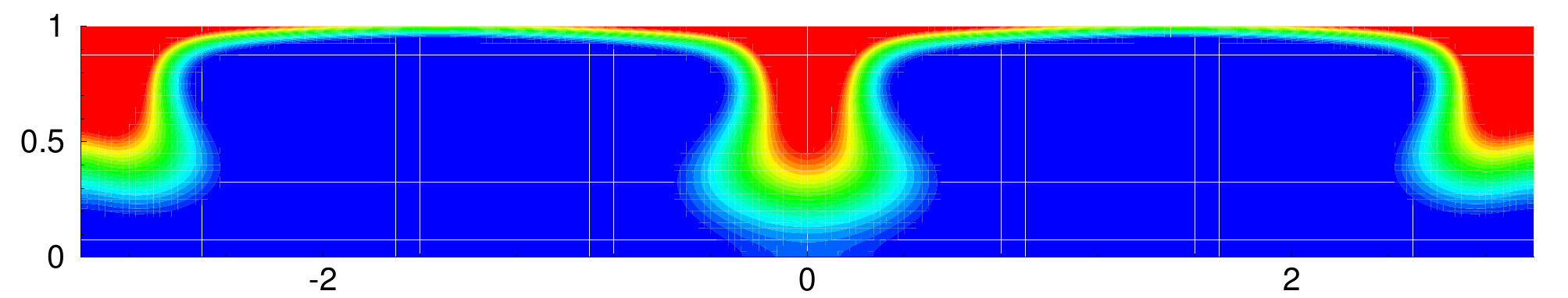}}}
\end{minipage}
\caption{Numerical results for $n$ with respect to the deterministic initial conditions. At the initial time, bacterial density is higher in the upper layer.
(a) initial condition is $n_0=[^{~1,~~ y \geq 0.5+0.1\cos({2\pi x}/{3})}
_{~0.5,~~ y < 0.5+0.1\cos({2\pi x}/{3})} $;
(b) simulation result at $t=1.2$ for the initial condition given in (a) ;
(c) initial condition is $n_0=[^{~1,~~ y \geq 0.5-0.1\cos({2\pi x}/{3})}
_{~0.5,~~ y < 0.5-0.1\cos({2\pi x}/{3})} $;
(d) simulation result at $t=1.2$ for the initial condition given in (c).}
\label{fig:fig1}
\end{figure}

\begin{figure}
\centering
\begin{minipage}{\linewidth}
\mbox{\subfigure[]{\includegraphics[width=0.45\linewidth]{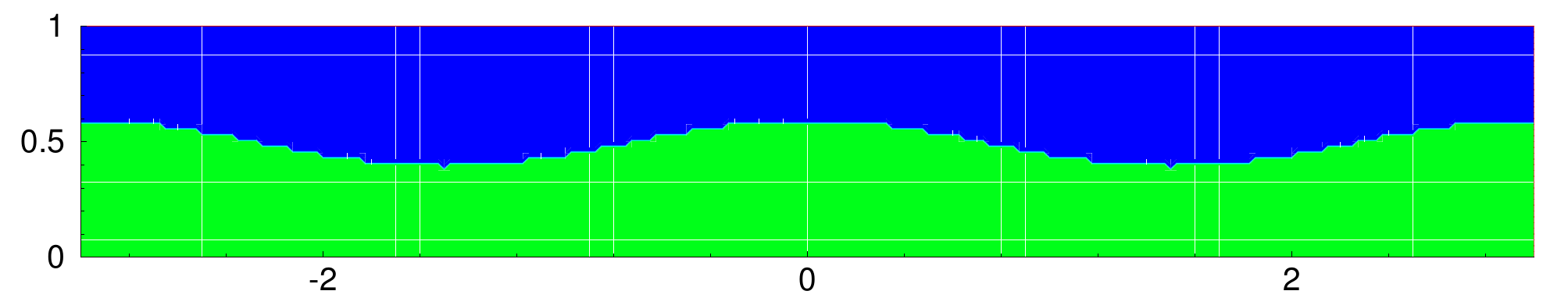}}
      \subfigure[]{\includegraphics[width=0.45\linewidth]{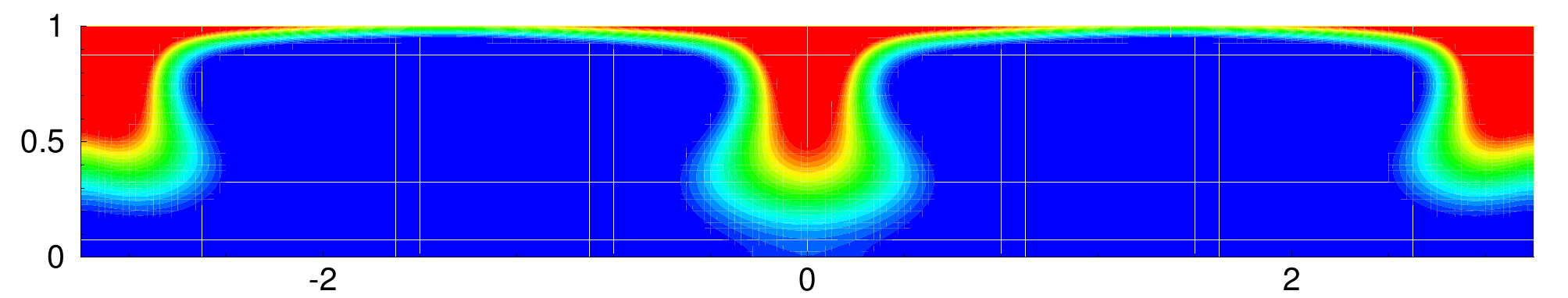}}}
\mbox{\subfigure[]{\includegraphics[width=0.45\linewidth]{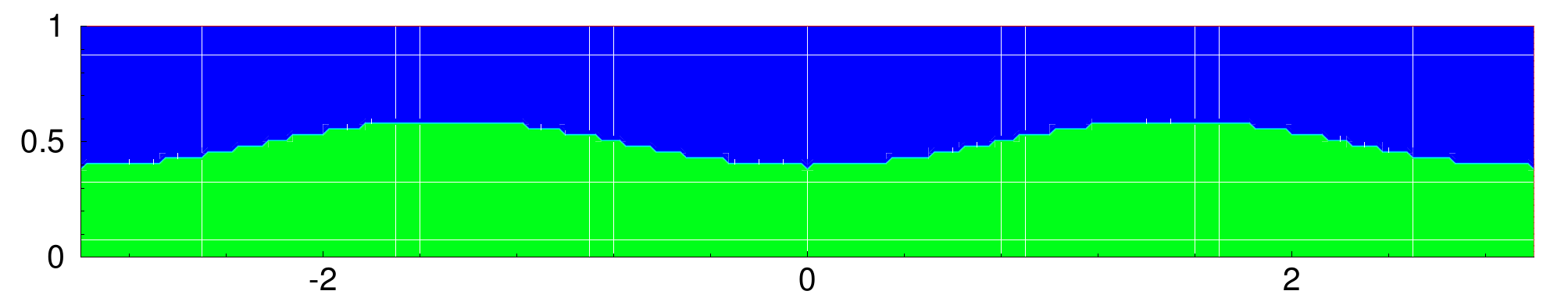}}
      \subfigure[]{\includegraphics[width=0.45\linewidth]{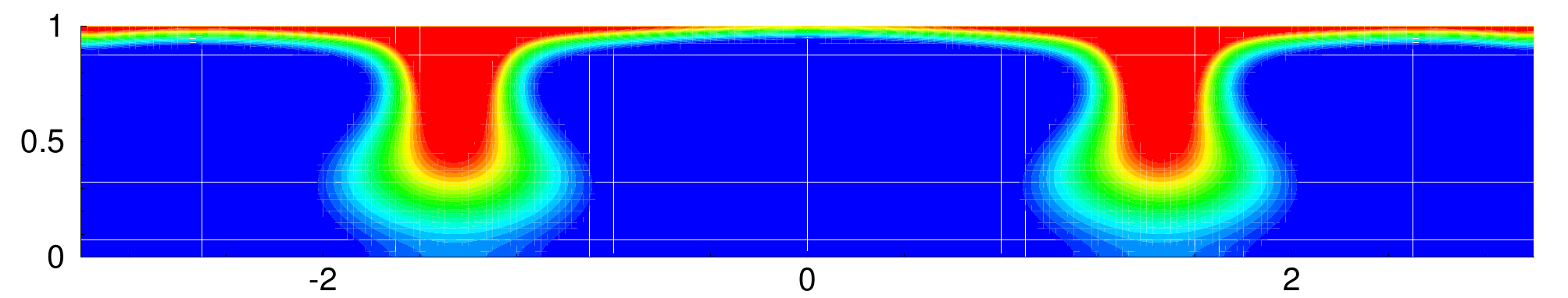}}}
\end{minipage}
\caption{Numerical results for $n$ with respect to the deterministic initial conditions. At the initial time, bacterial density is higher in the upper layer.
(a) initial condition is $n_0=[^{~0.5,~~ y \geq 0.5+0.1\cos({2\pi x}/{3})}
_{~1,~~ y < 0.5+0.1\cos({2\pi x}/{3})} $;
(b) simulation result at $t=1.2$ for the initial condition given in (a) ;
(c) initial condition is $n_0=[^{~0.5,~~ y \geq 0.5-0.1\cos({2\pi x}/{3})}
_{~1,~~ y < 0.5-0.1\cos({2\pi x}/{3})} $;
(d) simulation result at $t=1.2$ for the initial condition given in (c). }
\label{fig:fig2}
\end{figure}

All numerical results are computed at $\Lew_{\tau}=5$, $\Pran_{\tau}=500$, $\Ra_{\tau}=2000$, $\mathsf{S}=10$, and $\mathsf{H}=4$ to ensure the formation of bacterium rich plumes. We first consider a profile given by the form of wave function $\cos(2\pi x/3)$ to determine the two initial layers of bacteria. Simulation results and initial conditions are shown in figure \ref{fig:fig1}. The upper layer has a higher bacterial density than the lower layer. The plumes form at the initial location of the crest of the wave function where there is a larger number of bacteria.
By reversing the initial conditions with a denser inferior layer (figure \ref{fig:fig2}), the plumes also form at the same location. 
We can observe that distinct initial conditions can lead to the formation of a very similar pattern (figures~\ref{fig:fig1}-\ref{fig:fig2}).

%Fig06--------------------------------------------------------------------------------
%-------------------------------------------------------------------------------------

%Fig07--------------------------------------------------------------------------------
%-------------------------------------------------------------------------------------

\begin{figure}
\centering
\begin{minipage}{\linewidth}
\mbox{\subfigure[]{\includegraphics[width=0.45\linewidth]{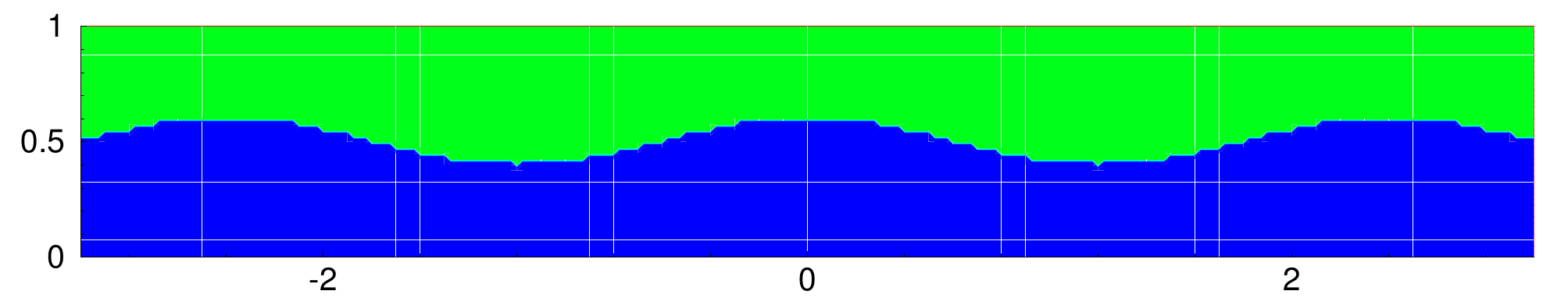}}
      \subfigure[]{\includegraphics[width=0.45\linewidth]{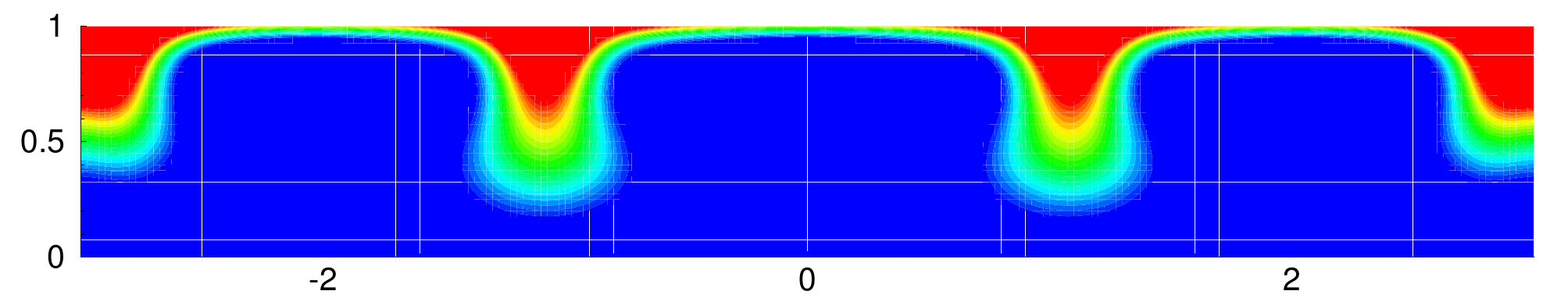}}}
\mbox{\subfigure[]{\includegraphics[width=0.45\linewidth]{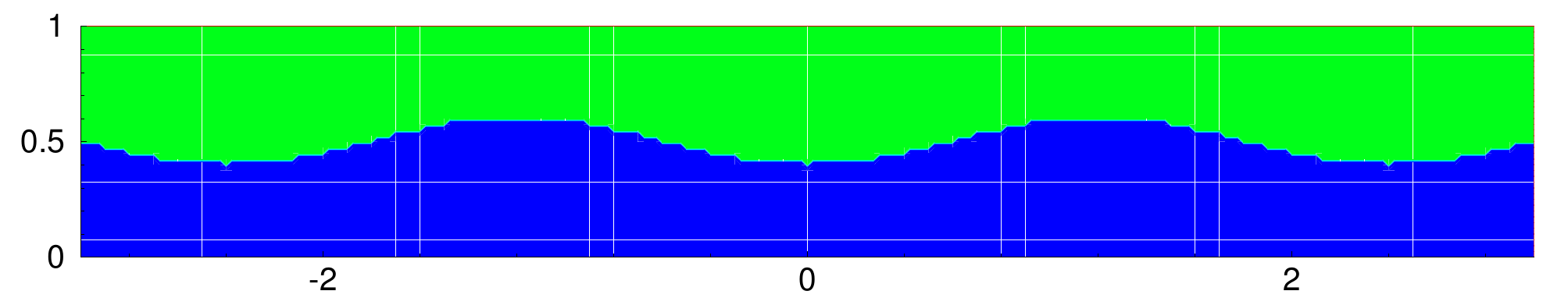}}
      \subfigure[]{\includegraphics[width=0.45\linewidth]{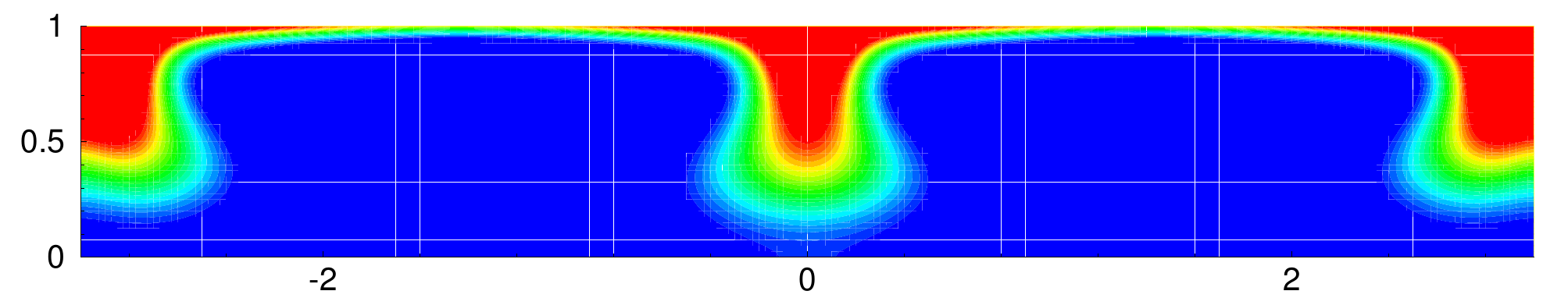}}}
\end{minipage}
\caption{Numerical results for $n$ with respect to the deterministic initial conditions. At the initial time, bacterial density is higher in the upper layer.
(a) initial condition is $n_0=[^{~1,~~ y \geq 0.5+0.1\cos({5\pi x}/{2})}
_{~0.5,~~ y < 0.5+0.1\cos({5\pi x}/{2})} $;
(b) simulation result at $t=1.2$ for the initial condition given in (a) ;
(c) initial condition is $n_0=[^{~1,~~ y \geq 0.5-0.1\cos({5\pi x}/{2})}
_{~0.5,~~ y < 0.5-0.1\cos({5\pi x}/{2})} $;
(d) simulation result at $t=1.2$ for the initial condition given in (c). }
\label{fig:fig3}
\end{figure}

\begin{figure}
\centering
\begin{minipage}{\linewidth}
\mbox{\subfigure[]{\includegraphics[width=0.45\linewidth]{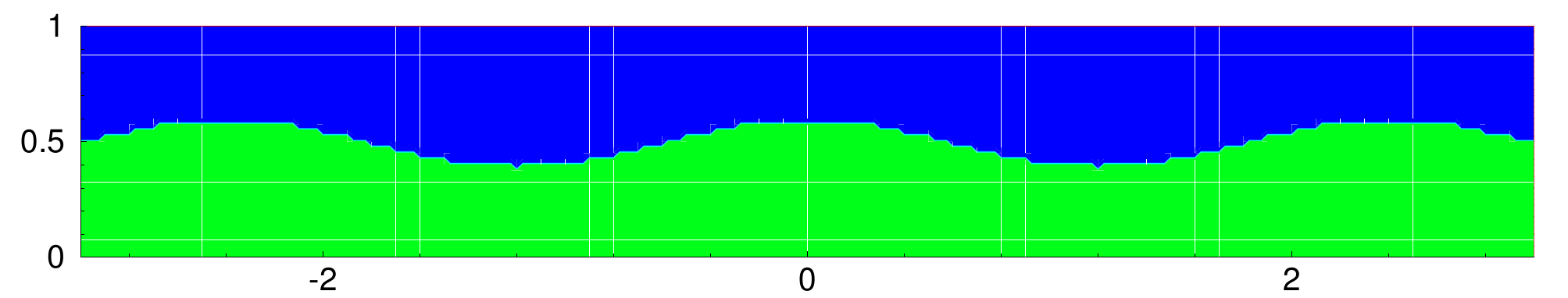}}
      \subfigure[]{\includegraphics[width=0.45\linewidth]{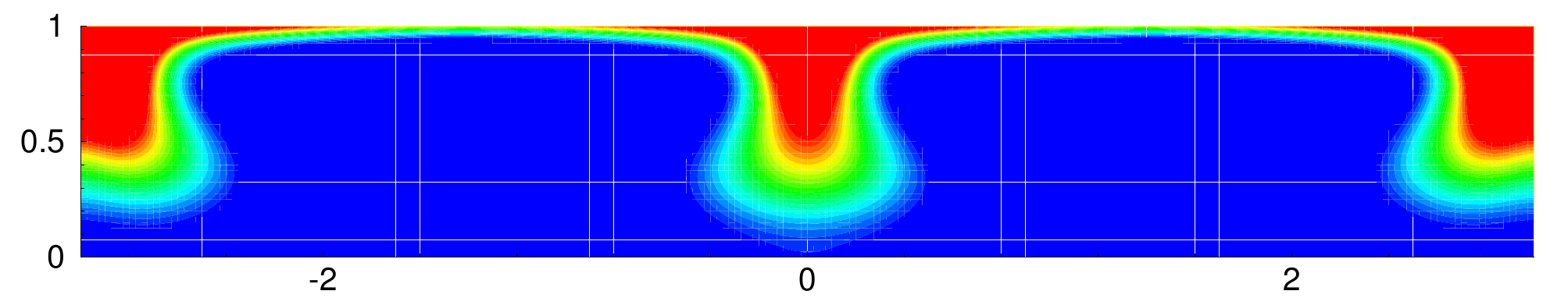}}}
\mbox{\subfigure[]{\includegraphics[width=0.45\linewidth]{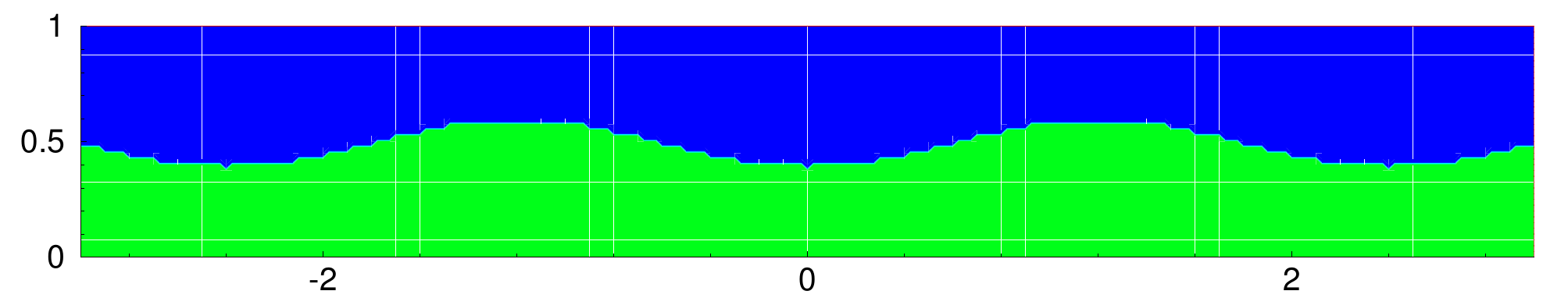}}
      \subfigure[]{\includegraphics[width=0.45\linewidth]{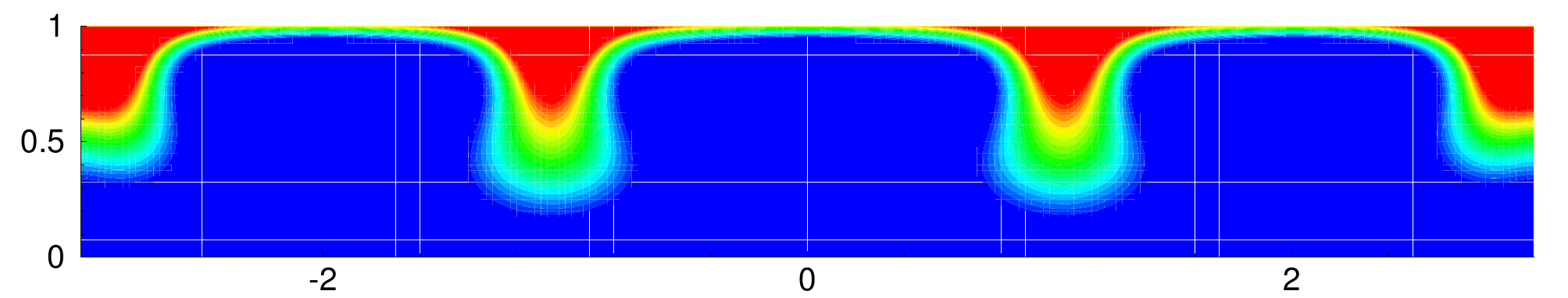}}}
\end{minipage}
\caption{Numerical results for $n$ with respect to the deterministic initial conditions. At the initial time, bacterial density is higher in the lower layer.
(a) initial condition is $n_0=[^{~0.5,~~ y \geq 0.5+0.1\cos({5\pi x}/{2})}
_{~1,~~ y < 0.5+0.1\cos({5\pi x}/{2})} $;
(b) simulation result at $t=1.2$ for the initial condition given in (a) ;
(c) initial condition is $n_0=[^{~0.5,~~ y \geq 0.5-0.1\cos({5\pi x}/{2})}
_{~1,~~ y < 0.5-0.1\cos({5\pi x}/{2})} $;
(d) simulation result at $t=1.2$ for the initial condition given in (c). }
\label{fig:fig4}
\end{figure}

We vary the profile of the curve between the upper and lower layers in the initial condition with different wave functions $\cos(3\pi x/2)$, $\cos(5\pi x/2)$, and $\cos(2\pi x)$. The wavenumber is defined as the spatial frequency of waves per unit distance. The wavelength is defined as the domain horizontal length divided by the wavenumber. Increasing the wavenumber of the initial wave function causes the formation of additional plumes to occur (figures~\ref{fig:fig3} and~\ref{fig:fig4}). In all simualtions, the locations of the plumes correspond to sites of initially greater local bacterial density.

%Fig08--------------------------------------------------------------------------------
%-------------------------------------------------------------------------------------

%Fig09--------------------------------------------------------------------------------
%-------------------------------------------------------------------------------------

\begin{figure}
\centering
\begin{minipage}{\linewidth}
\mbox{\subfigure[]{\includegraphics[width=0.45\linewidth]{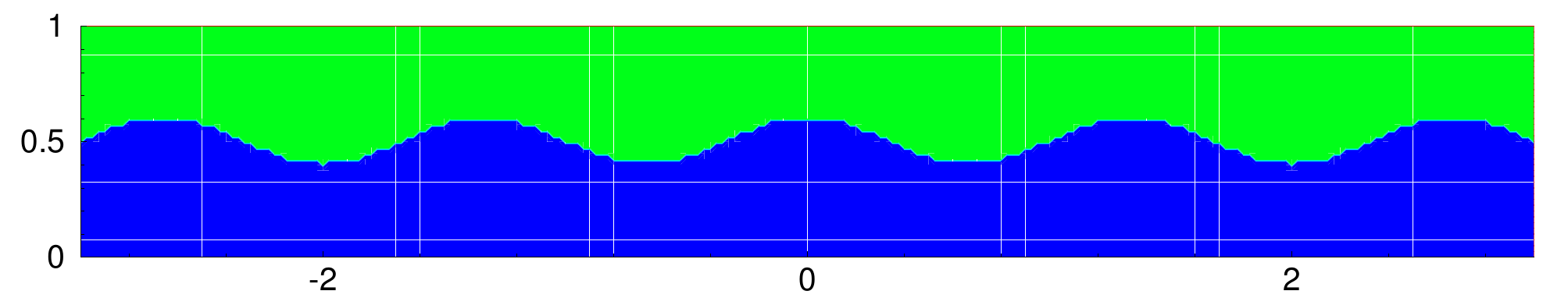}}
      \subfigure[]{\includegraphics[width=0.45\linewidth]{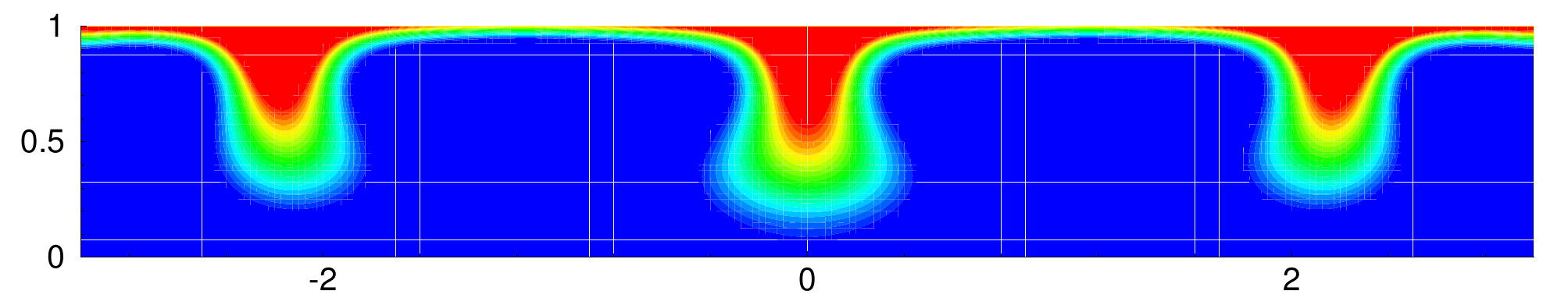}}}
\mbox{\subfigure[]{\includegraphics[width=0.45\linewidth]{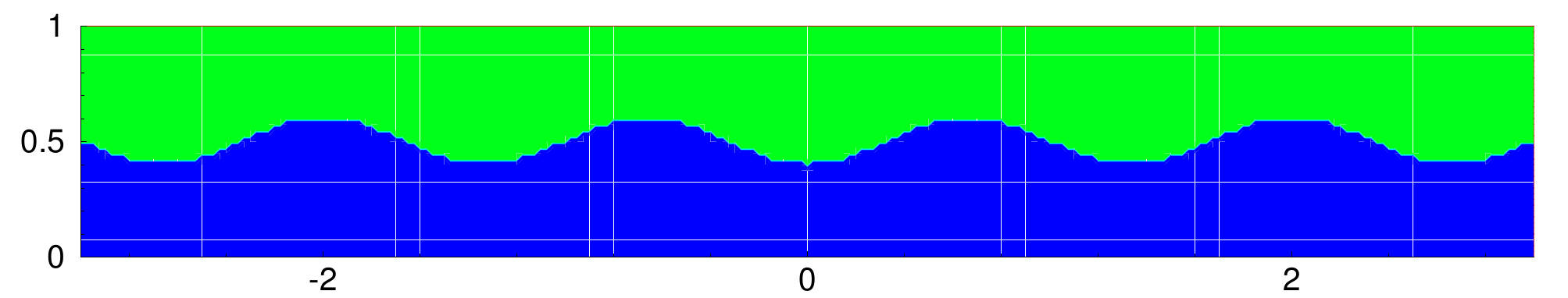}}
      \subfigure[]{\includegraphics[width=0.45\linewidth]{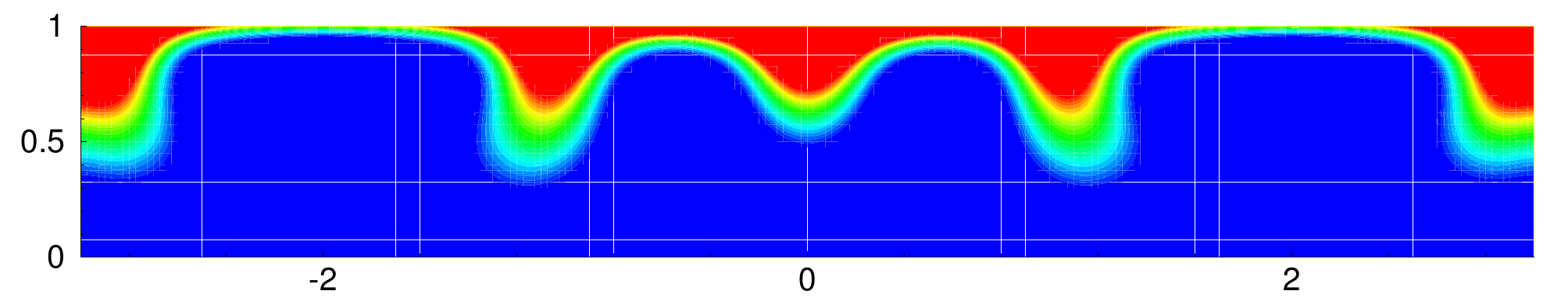}}}
\end{minipage}
\caption{Numerical results for $n$ with respect to the deterministic initial conditions. At the initial time, bacterial density is higher in the upper layer.
(a) initial condition is $n_0=[^{~1,~~ y \geq 0.5+0.1\cos({3\pi x}/{2})}
_{~0.5,~~ y < 0.5+0.1\cos({3\pi x}/{2})} $;
(b) simulation result at $t=1.2$ for the initial condition given in (a) ;
(c) initial condition is $n_0=[^{~1,~~ y \geq 0.5-0.1\cos({3\pi x}/{2})}
_{~0.5,~~ y < 0.5-0.1\cos({3\pi x}/{2})} $;
(d) simulation result at $t=1.2$ for the initial condition given in (c). }
\label{fig:fig5}
\end{figure}

\begin{figure}
\centering
\begin{minipage}{\linewidth}
\mbox{\subfigure[]{\includegraphics[width=0.45\linewidth]{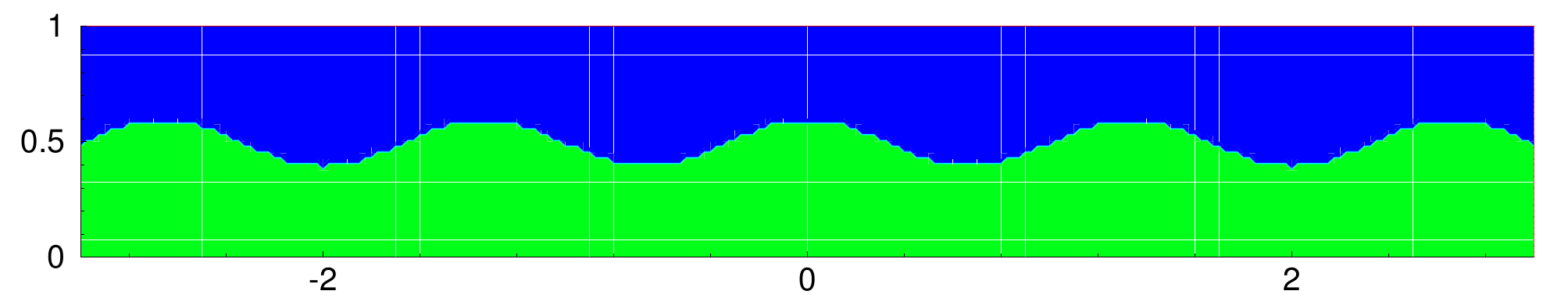}}
      \subfigure[]{\includegraphics[width=0.45\linewidth]{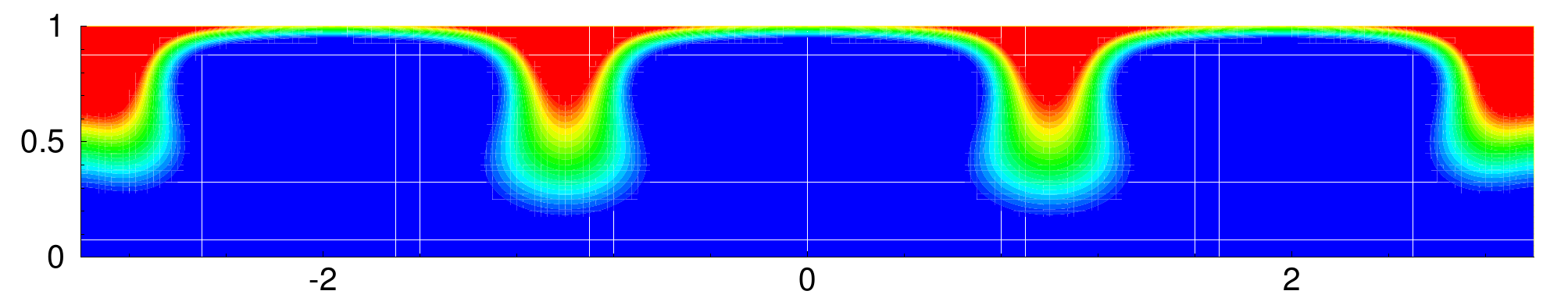}}}
\mbox{\subfigure[]{\includegraphics[width=0.45\linewidth]{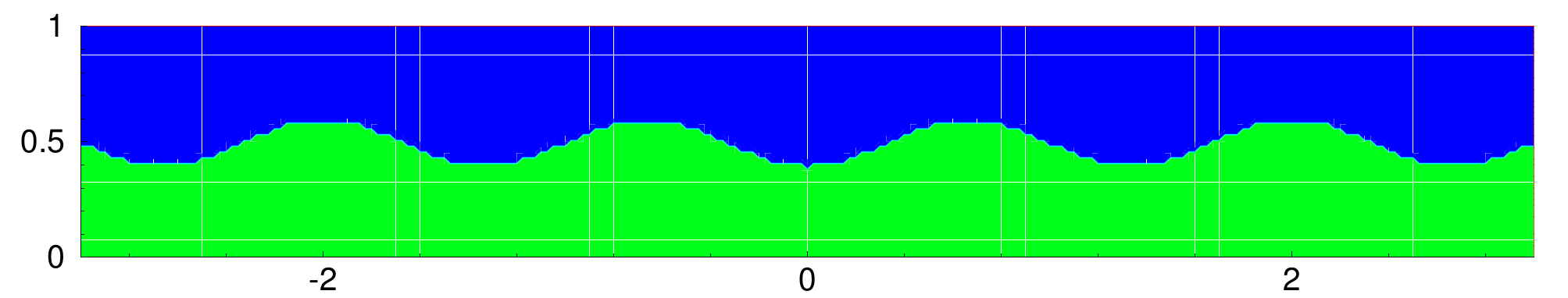}}
      \subfigure[]{\includegraphics[width=0.45\linewidth]{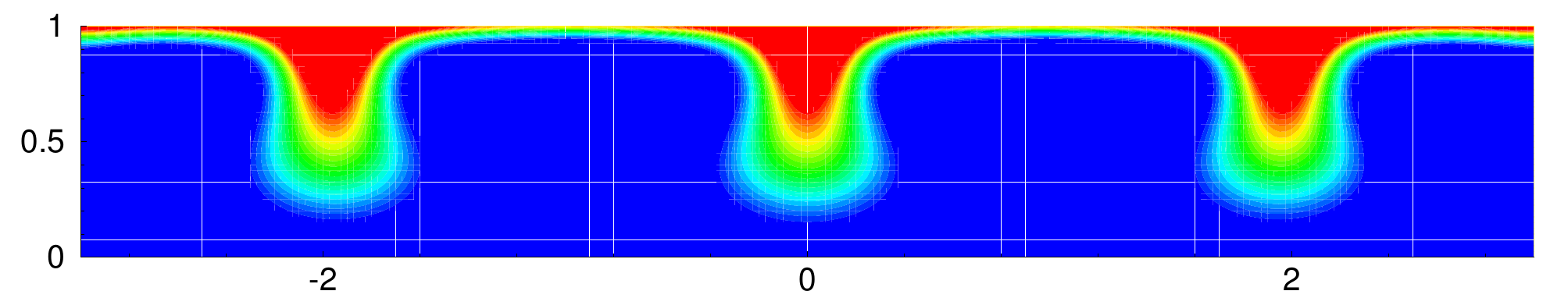}}}
\end{minipage}
\caption{Numerical results for $n$ with respect to the deterministic initial conditions. At the initial time, bacterial density is higher in the lower layer.
(a) initial condition is $n_0=[^{~0.5,~~ y \geq 0.5+0.1\cos({3\pi x}/{2})}
_{~1,~~ y < 0.5+0.1\cos({3\pi x}/{2})} $;
(b) simulation result at $t=1.2$ for the initial condition given in (a) ;
(c) initial condition is $n_0=[^{~0.5,~~ y \geq 0.5-0.1\cos({3\pi x}/{2})}
_{~1,~~ y < 0.5-0.1\cos({3\pi x}/{2})} $;
(d) simulation result at $t=1.2$ for the initial condition given in (c). }
\label{fig:fig6}
\end{figure}

\begin{figure}
\centering
\begin{minipage}{\linewidth}
\mbox{\subfigure[]{\includegraphics[width=0.45\linewidth]{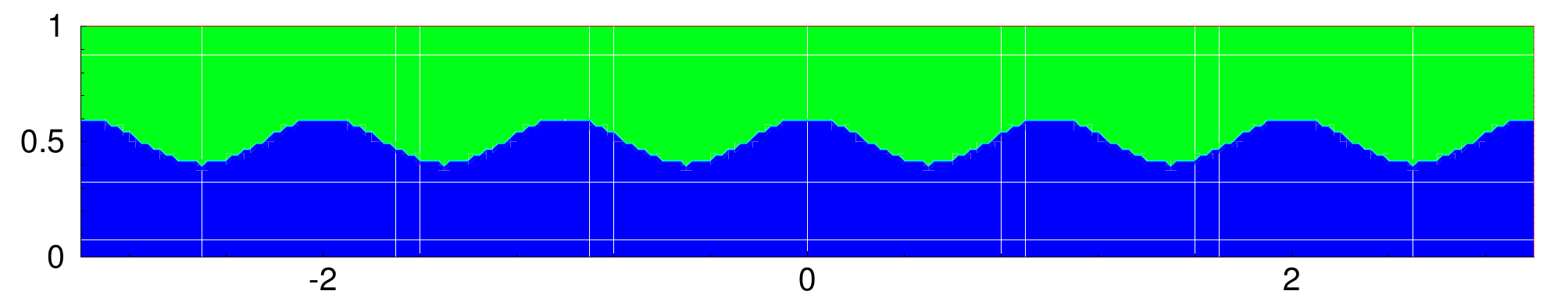}}
      \subfigure[]{\includegraphics[width=0.45\linewidth]{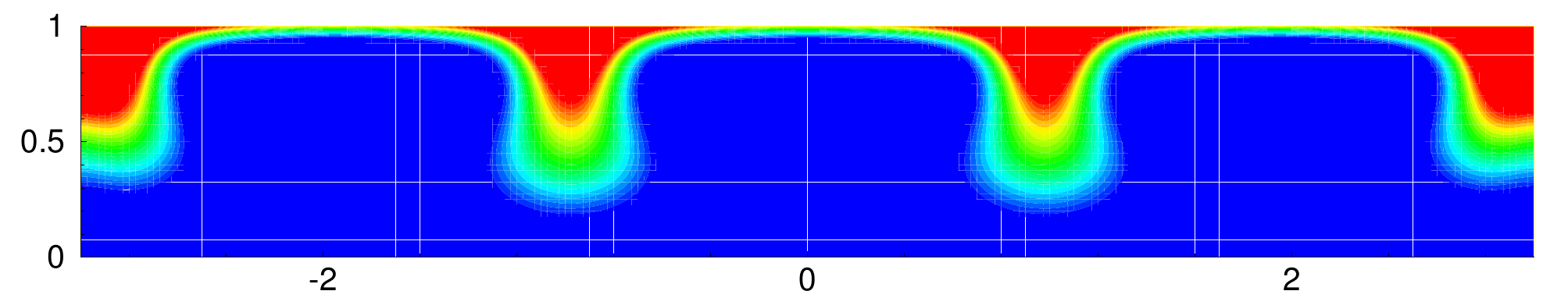}}}
\mbox{\subfigure[]{\includegraphics[width=0.45\linewidth]{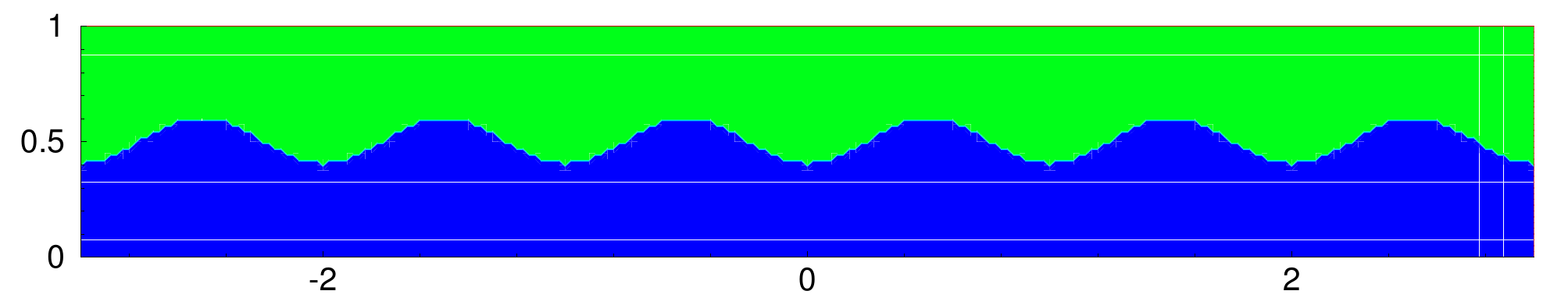}}
      \subfigure[]{\includegraphics[width=0.45\linewidth]{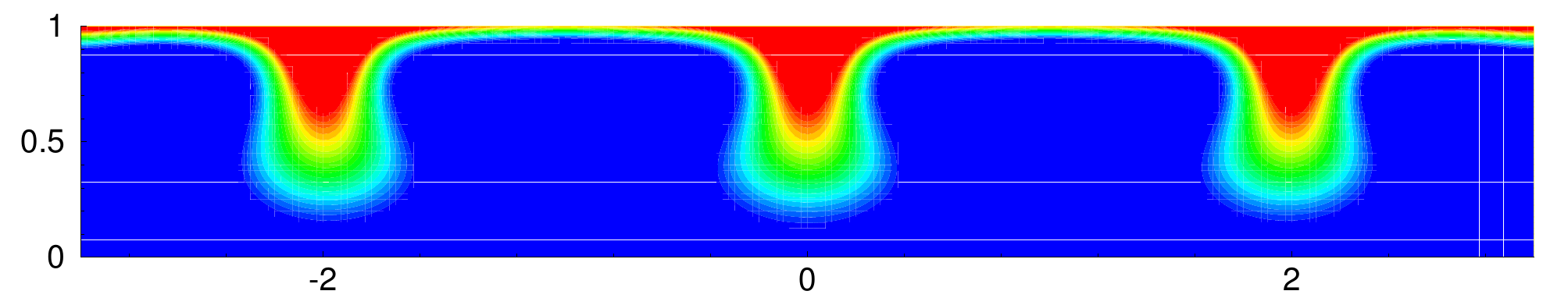}}}
\end{minipage}
\caption{Numerical results for $n$ with respect to the deterministic initial conditions. At the initial time, bacterial density is higher in the upper layer.
(a) initial condition is $n_0=[^{~1,~~ y \geq 0.5+0.1\cos(2\pi x)}
_{~0.5,~~ y < 0.5+0.1\cos(2\pi x)} $;
(b) simulation result at $t=1.2$ for the initial condition given in (a) ;
(c) initial condition is $n_0=[^{~1,~~ y \geq 0.5-0.1\cos(2\pi x)}
_{~0.5,~~ y < 0.5-0.1\cos(2\pi x)} $;
(d) simulation result at $t=1.2$ for the initial condition given in (c). }
\label{fig:fig7}
\end{figure}

\begin{figure}
\centering
\begin{minipage}{\linewidth}
\mbox{\subfigure[]{\includegraphics[width=0.45\linewidth]{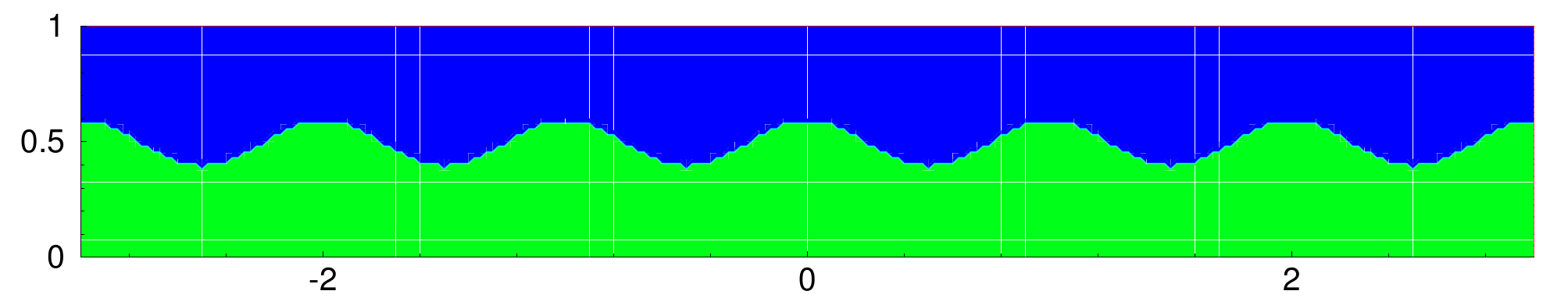}}
      \subfigure[]{\includegraphics[width=0.45\linewidth]{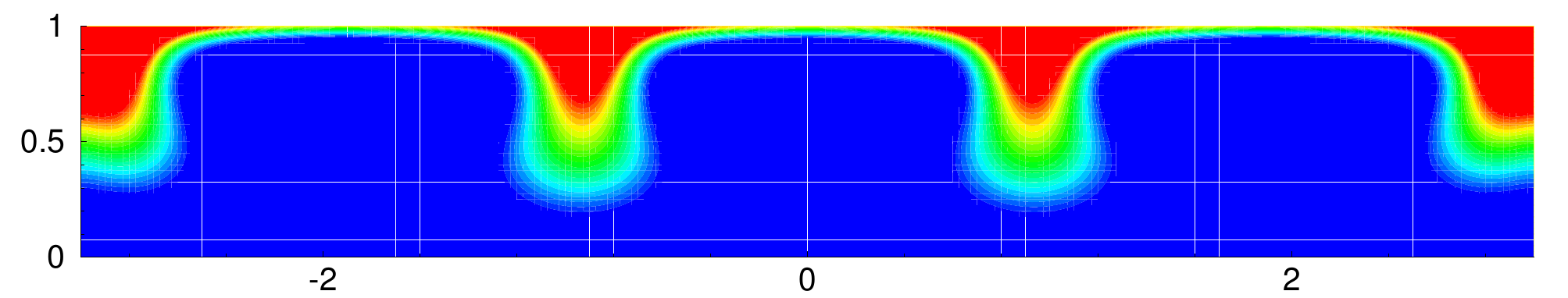}}}
\mbox{\subfigure[]{\includegraphics[width=0.45\linewidth]{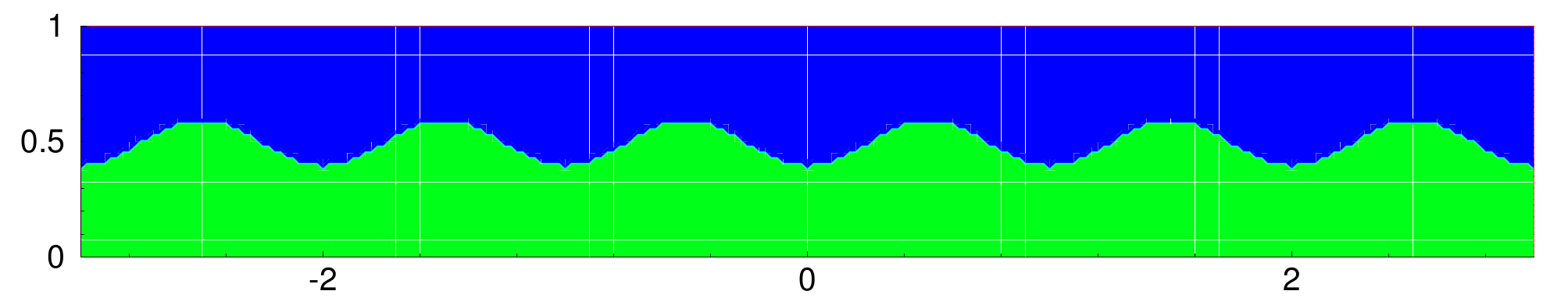}}
      \subfigure[]{\includegraphics[width=0.45\linewidth]{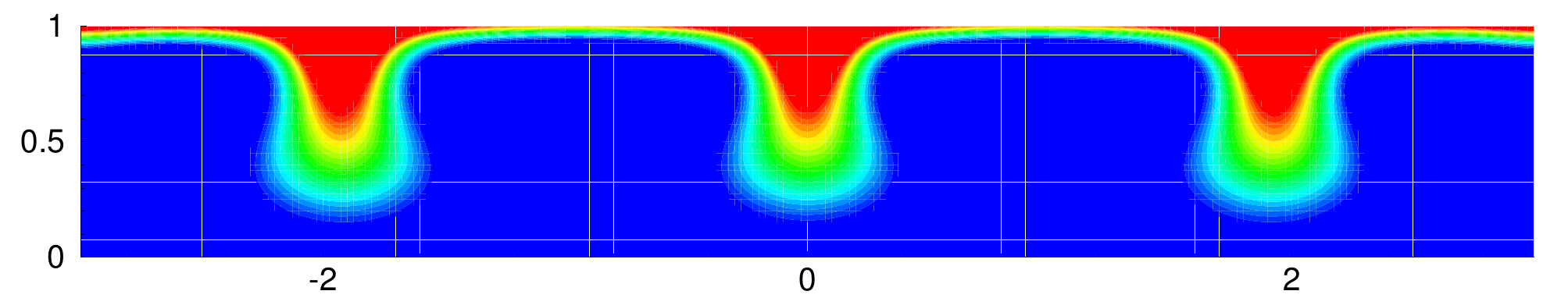}}}
\end{minipage}
\caption{Numerical results for $n$ with respect to the deterministic initial conditions. At the initial time, bacterial density is higher in the lower layer.
(a) initial condition is $n_0=[^{~0.5,~~ y \geq 0.5+0.1\cos(2\pi x)}
_{~1,~~ y < 0.5+0.1\cos(2\pi x)} $;
(b) simulation result at $t=1.2$ for the initial condition given in (a) ;
(c) initial condition is $n_0=[^{~0.5,~~ y \geq 0.5-0.1\cos(2\pi x)}
_{~1,~~ y < 0.5-0.1\cos(2\pi x)} $;
(d) simulation result at $t=1.2$ for the initial condition given in (c). }
\label{fig:fig8}
\end{figure}

When the wavenumber in the initial profile is large enough, the number of plumes is not equal to the wavenumber fixed by the initial condition. Three plumes, at most, form (figures~\ref{fig:fig5} to~\ref{fig:fig8}). Nonetheless, more than three plumes can form, but they merge later (figure~\ref{fig:fig7}). Plume merging was previously observed in \cite{chertock_sinking_2012}. The plume merging mechanism is analogous to that of the Rayleigh-B\'enard flow. Spacing between plumes seems to be intrinsic to the system and the position can only be predicted in very specific cases such as the examples given above.

%Fig10--------------------------------------------------------------------------------
%-------------------------------------------------------------------------------------

%Fig11--------------------------------------------------------------------------------
%-------------------------------------------------------------------------------------

%Fig12--------------------------------------------------------------------------------
%-------------------------------------------------------------------------------------

%Fig13--------------------------------------------------------------------------------
%-------------------------------------------------------------------------------------

\begin{figure}
\centering
\begin{minipage}{\linewidth}
\mbox{\subfigure[]{\includegraphics[width=0.45\linewidth]{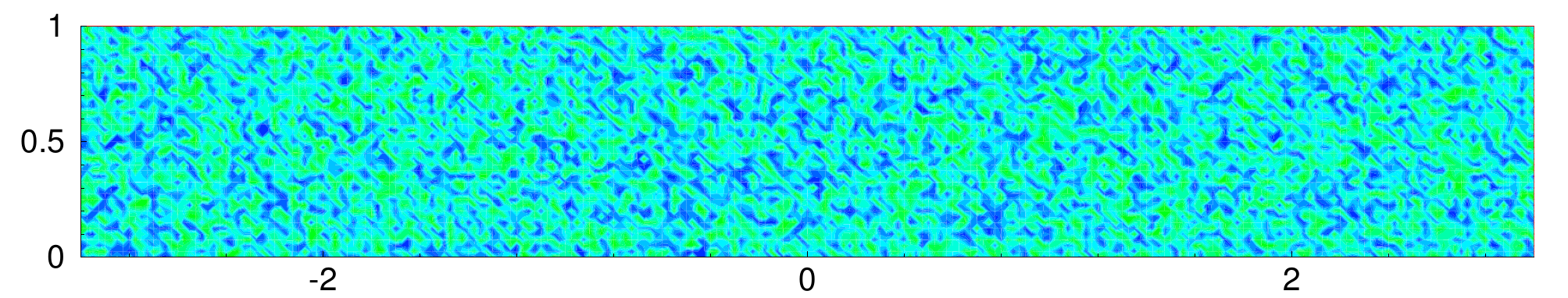}}
      \subfigure[]{\includegraphics[width=0.45\linewidth]{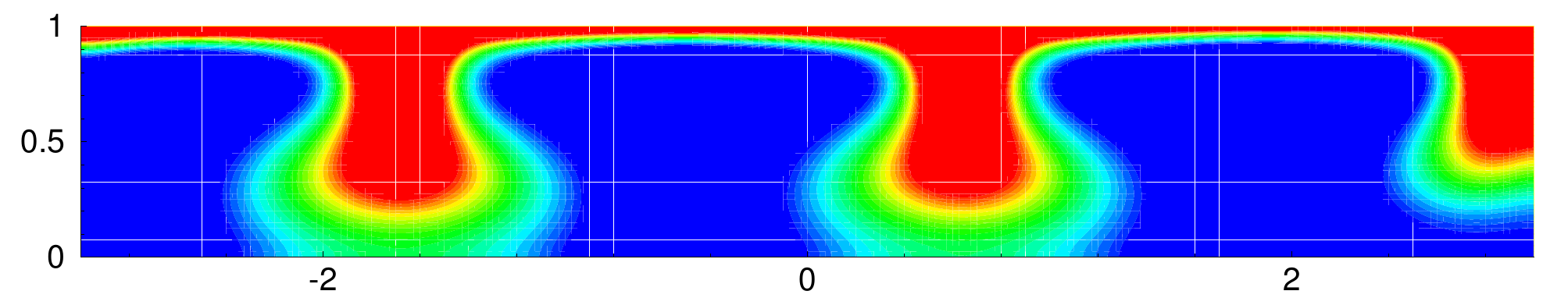}}}
\mbox{\subfigure[]{\includegraphics[width=0.45\linewidth]{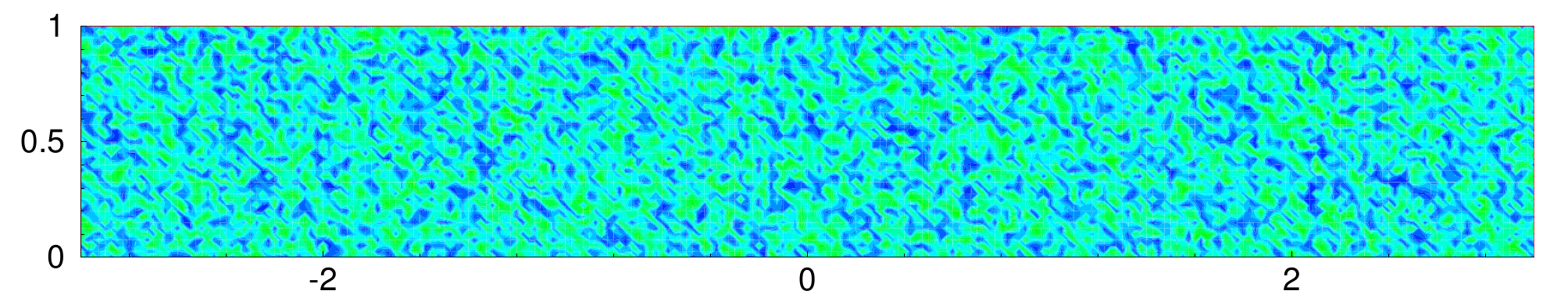}}
      \subfigure[]{\includegraphics[width=0.45\linewidth]{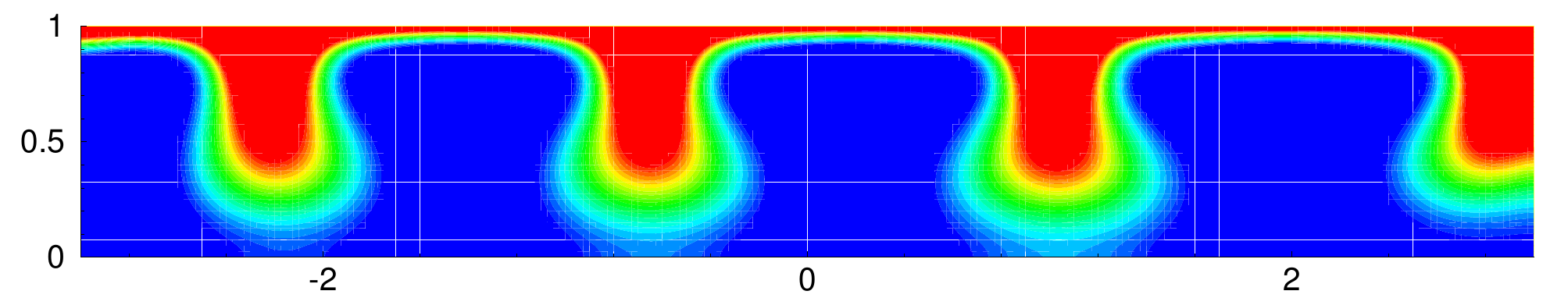}}}
\mbox{\subfigure[]{\includegraphics[width=0.45\linewidth]{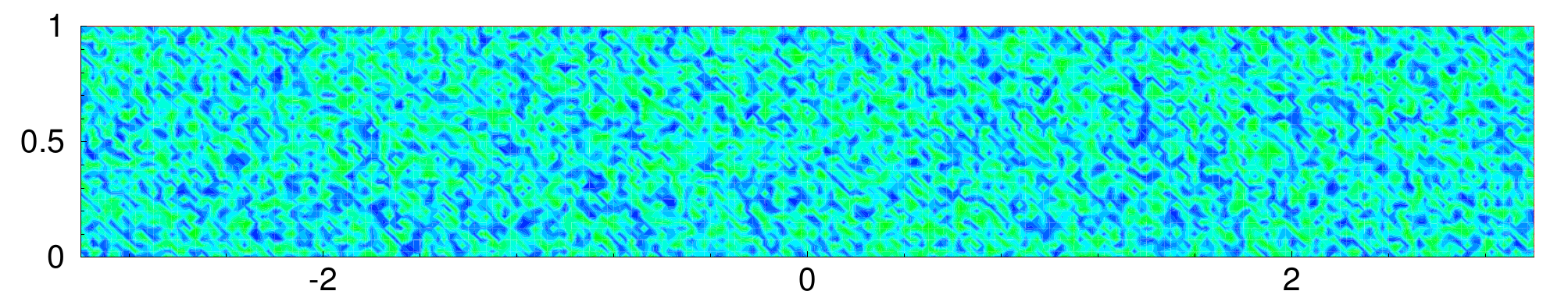}}
      \subfigure[]{\includegraphics[width=0.45\linewidth]{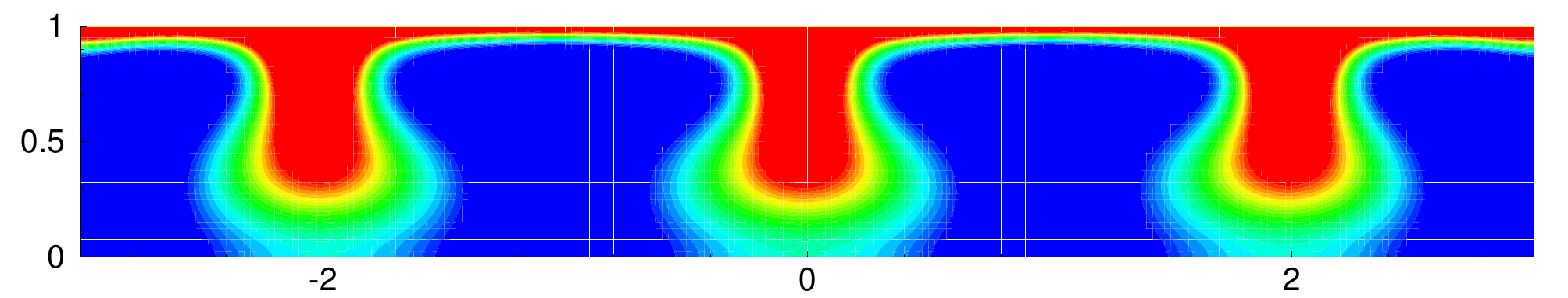}}}
\end{minipage}
\caption{Simulations subject to the initial condition given in (\ref{randomIC}) (left) and the corresponding numerical results for $n$ at $t=1.2$ (right).}
\label{fig:fig9}
\end{figure}

The randomly perturbed bacterial density is defined as follows \cite{chertock_sinking_2012}:
\beqt
n(\mathrm{x})=0.8 + 0.2\varepsilon(\mathrm{x}) ,
\label{randomIC}
\eeqt
with $\varepsilon$ being a random number with a uniform probability distribution over $[0,1]$. Simulation results with the randomly perturbed initial condition are given in figure~\ref{fig:fig9}. We note that the plume-to-plume spacing is not fixed but the number of plumes in the solution remains the same as that with the solution obtained subject to the deterministic initial condition illustrated in figures \ref{fig:fig5} to \ref{fig:fig8}.

%Fig14--------------------------------------------------------------------------------
%-------------------------------------------------------------------------------------

In figures \ref{fig:fig1}--\ref{fig:fig9}, we can see plumes forming at the border of the domain. These border plumes seem to be caused by the no-slip boundary condition on the velocity depending on the arrangement of the convection cells. Bacteria first agglomerate at the surface near the wall. Because the fluid velocity is equal to zero at the wall boundary, a large amount of bacteria remains close to the wall. At a location away from the wall, the velocity is non-zero, thus bacteria descend into the fluid and form a plume at the border.

The location of the plumes may only be predicted in very simple tests for which the wavelengths of initial conditions are lower than the wavelength of the process. Despite the difficulty to predict the site of plume formation, the system presents a dominant wavelength of the fingering instability. The wavelength and wavenumber are redefined such as those given in \cite{pan_modelling_2001}. The wavelength is the length of the domain divided by the number of plumes and the wavenumber $2\pi$ divided by the wavelength. The wavelength and wavenumber are similar for all simulation tests whatever the length of the domain is (tables~\ref{tab3:l2} to~\ref{tab6:l5}). 

%Table3--------------------------------------------------------------------------------
\begin{table}
\caption{The predicted number of plumes, wavenumber and wavelength for $\ell=2$. Each row corresponds to a different simulation result subject to the randomly perturbed initial condition given in (\ref{randomIC}). }
\label{tab3:l2}
\centering
\begin{tabular}{ccc}
number of plumes    & wavelength & wavenumber\\ \hline
2     & 2      & 3.14     \\
2     & 2      & 3.14     \\
2     & 2      & 3.14     \\
2     & 2      & 3.14     \\
\end{tabular}
\end{table}
%-------------------------------------------------------------------------------------

%Table4--------------------------------------------------------------------------------
\begin{table}
\caption{The predicted number of plumes, wavenumber and wavelength for $\ell=3$. Each simulation result is subject to the random initial condition given in (\ref{randomIC}). }
\label{tab4:l3}
\centering
\begin{tabular}{ccc}
number of plumes    & wavelength & wavenumber\\ \hline
3     & 2      & 3.14     \\
3     & 2      & 3.14     \\
4     & 1.5    & 4.19     \\
3     & 2      & 3.14     \\
\end{tabular}
\end{table}
%-------------------------------------------------------------------------------------

%Table5--------------------------------------------------------------------------------
\begin{table}
\caption{The predicted number of plumes, wavenumber and wavelength for $\ell=4$. Each simulation result is subject to the random initial condition given in (\ref{randomIC}). }
\label{tab5:l4}
\centering

\begin{tabular}{ccc}
number of plumes    & wavelength & wavenumber\\ \hline
5     & 1.6      & 3.93     \\
5     & 1.6      & 3.93     \\
5     & 1.6      & 3.93     \\
5     & 1.6      & 3.93     \\
%Average & 1.6    & 3.93     \\
\end{tabular}
\end{table}
%-------------------------------------------------------------------------------------

%Table6--------------------------------------------------------------------------------
\begin{table}
\caption{The predicted number of plumes, wavenumber and wavelength for $\ell=5$. Each simulation result is subject to the random initial condition given in (\ref{randomIC}). }
\label{tab6:l5}
\centering
\begin{tabular}{ccc}
number of plumes    & wavelength & wavenumber\\ \hline
5     & 2        & 3.14     \\
6     & 1.67     & 3.77     \\
5     & 2        & 3.14     \\
5     & 2        & 3.14     \\
%Average & 1.92    & 3.28     \\
\end{tabular}
\end{table}
%-------------------------------------------------------------------------------------

\subsubsection{Growth of plumes}
\label{growthplumes}

We now arbitrarily define the plumes by the isoline $n=1$. This choice allows us to track the stack layer and the plumes being formed. The layer where $n>1$ represents the layer with a high bacterial density from which plumes form and descend in the fluid. 
%The condition $n>1$ indicates that the bacterial density is larger than the initial average density $\overline{n_0}$ defined in (\ref{n0}).
On the other hand, the layer with $n<1$ corresponds to the depletion layer, from which bacteria have migrated to the stack layer.

\begin{figure}
\centering
\subfigure[]{\includegraphics[width=0.6\linewidth]{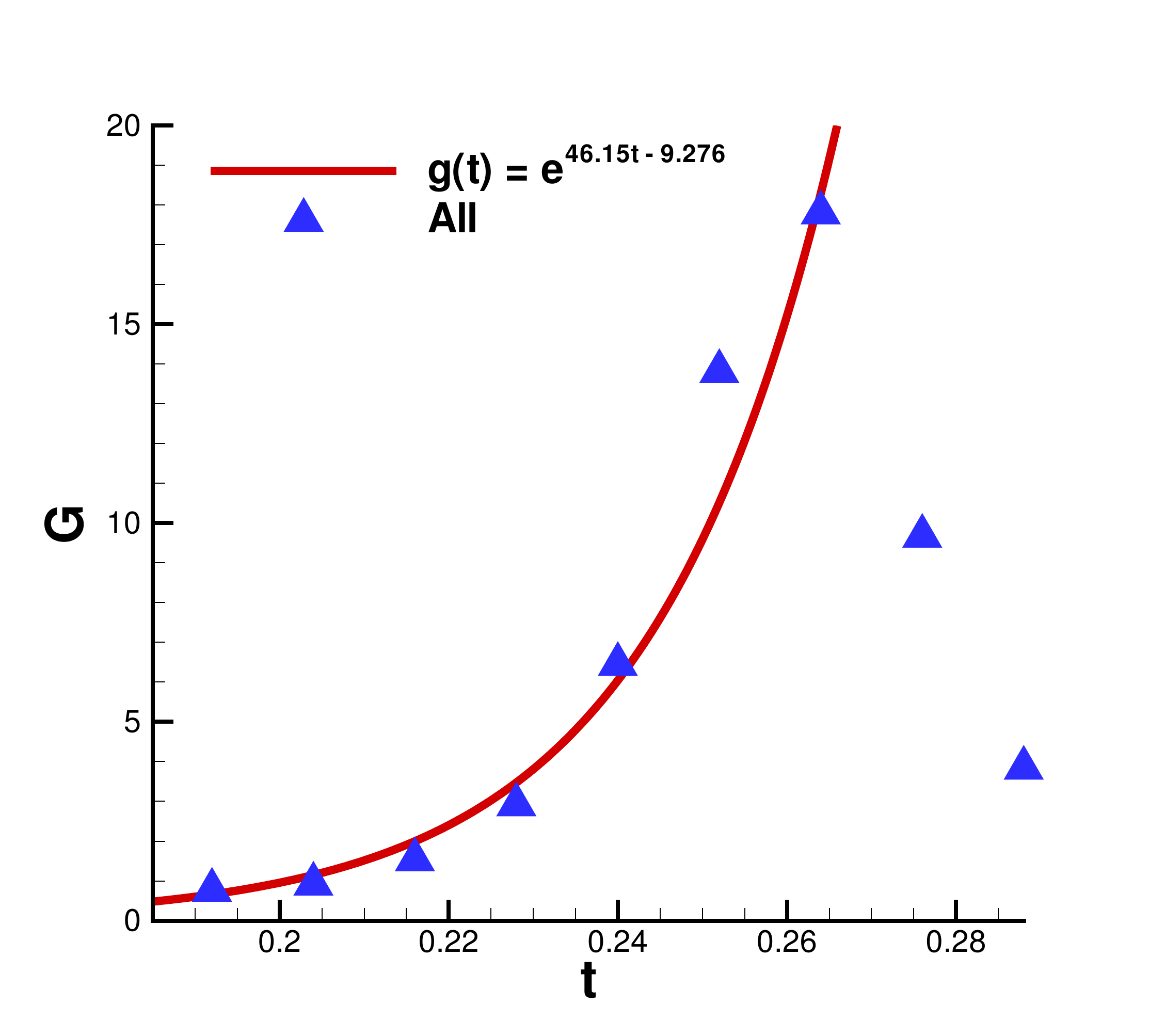}}
\subfigure[]{\includegraphics[width=0.6\linewidth]{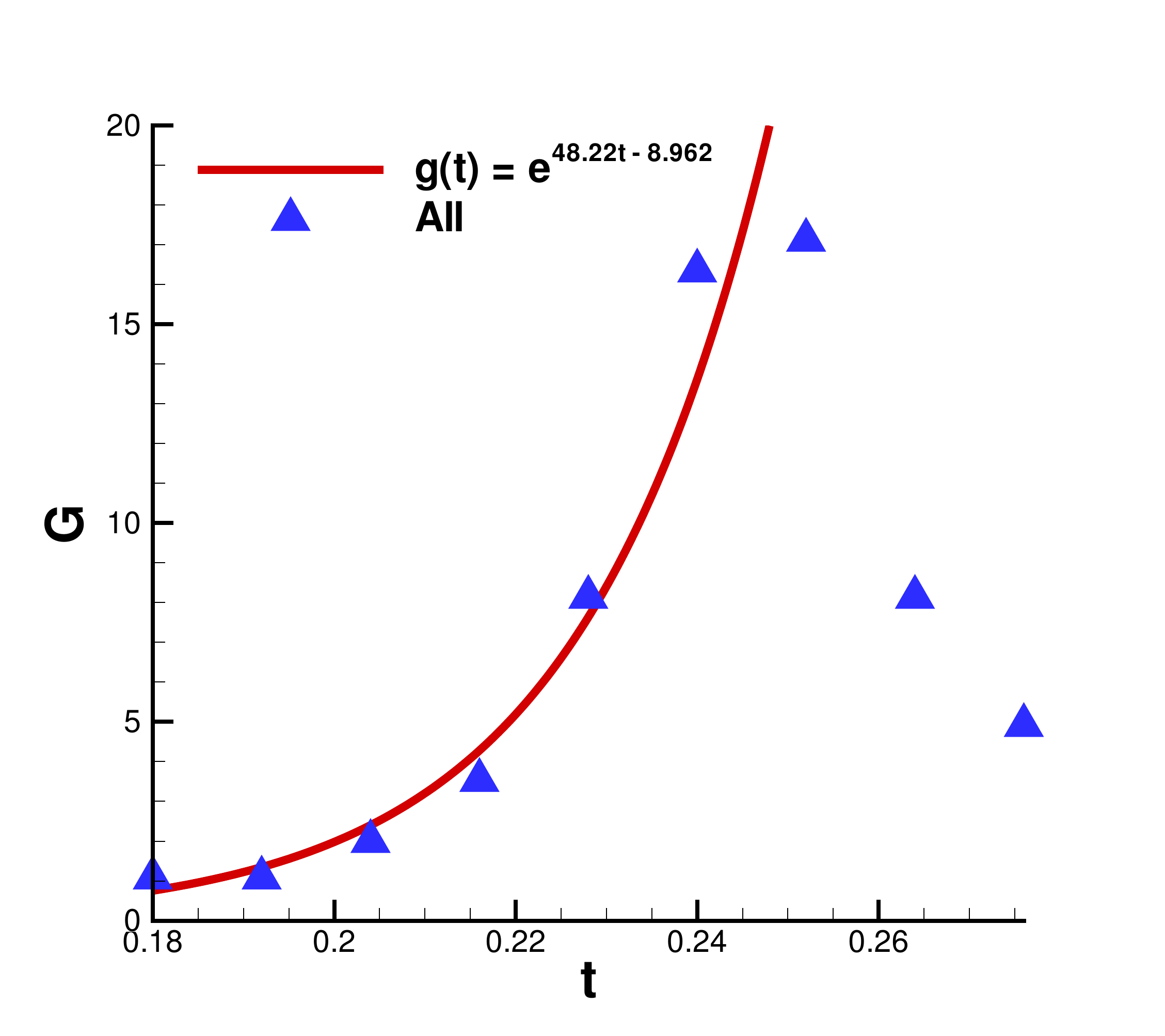}}
\caption{Plot of the growth rate $\mathsf{G}$ of plume amplitude (dots) corresponding to the data in table \ref{tab7:plumesgrowth}. The growth rate is interpolated by the exponential function $g(t)=e^{\alpha t + \beta}$ (line).}
\label{fig:fig19}
\end{figure}

We define the growth rate of the plume amplitudes by $\mathsf{G}=(A_t-A_{t-1})/\Delta t$. The plume amplitude $A_t$ is computed by measuring the distance from the surface to the tip of the plume at time $t$ (table~\ref{tab7:plumesgrowth}) and $\Delta t$ is the time increment. Data obtained can be interpolated by the exponential function $g(t)=\exp\{\alpha t + \beta\}$ as the amplitudes of the descending plumes grow exponentially (figure~\ref{fig:fig19}). The exponential growth is well understood in the conventional Rayleigh-Taylor convection. Subsequent to the exponential growth phase, the growth rate of amplitude decreases, as plumes get closer to the bottom of the container.

%Fig15--------------------------------------------------------------------------------
%-------------------------------------------------------------------------------------

%Table7--------------------------------------------------------------------------------
\begin{table}
\caption{The predicted growth rate $\mathsf{G}$ of the plume amplitudes $A_t$ for two simulations subject to the random initial condition (\ref{randomIC}).}
\label{tab7:plumesgrowth}
\centering
\begin{tabular}{ccccccc}
\multicolumn{3}{c}{(a)} & &\multicolumn{3}{c}{(b)} \\
\cmidrule(r){1-3}\cmidrule(r){5-7}
$A_t$    & $t$ & $\mathsf{G}$  & &$A_t$    & $t$ & $\mathsf{G}$ \\
0.18 & 0.16 & ~     & & 0.18  & 0.16 & ~        \\
0.19 & 0.18 & 0.46  & & 0.20  & 0.18 & 1.07     \\
0.20 & 0.19 & 0.76  & & 0.21  & 0.19 & 1.07     \\
0.21 & 0.20 & 0.92  & & 0.23  & 0.20 & 1.99     \\
0.23 & 0.21 & 1.53  & & 0.28  & 0.21 & 3.53     \\
0.26 & 0.22 & 2.91  & & 0.37  & 0.22 & 8.13     \\
0.34 & 0.24 & 6.44  & & 0.57  & 0.24 & 16.34    \\
0.51 & 0.25 & 13.81 & & 0.78  & 0.25 & 17.12    \\
0.72 & 0.26 & 17.81 & & 0.87  & 0.26 & 8.13     \\
0.83 & 0.27 & 9.67  & & 0.93  & 0.27 & 4.91     
\end{tabular}
\end{table}
%-------------------------------------------------------------------------------------

\subsection{Comparison with other buoyancy-driven convections}
\label{buoyancyconvections}

The chemotaxis--diffusion--convection system has many features similar to other well known buoyancy-driven flows, such as the double diffusive and Rayleigh B\'enard convection.

Double diffusive convection occurs in a fluid containing at least two components with different diffusivities. A destabilizing component diffuses faster than the stabilizing one \cite{lemaigre_asymmetric_2013}. The distinct diffusivities yield a density difference capable of driving the motion of fluid \cite{turner_double-diffusive_1974}.

Comparison with a classical example of double diffusive phenomenon in oceanography can be carried out by considering two superposed fluid layers with a specific combination of temperature ($T$) and the solute concentration ($s$), namely the salinity. The system of dimensionless equations of the double diffuse problem is the following:
\beqt
\label{eqDDC}
\begin{aligned}
  \frac{\partial \mathbf{u}}{\partial t}  {+} (\mathbf{u} \bcdot \bnabla)\mathbf{u}  {-}  \Pran_T \nabla^2 \mathbf{u} {+} \Pran_T \bnabla p & =  {-}  \Pran_T \, ( \Ra_m \, s {-} \Ra_T T) {\mathbf j} , \\
 \bnabla \bcdot \mathbf{u} & = 0, \\
  \frac{\partial T}{\partial t} +\mathbf{u} \bcdot \bnabla T  -   \nabla^2 T & =  0 , \\
    \frac{\partial s}{\partial t} +\mathbf{u} \bcdot \bnabla s  -  \Lew_T \, \nabla^2 s & =  0,
 \end{aligned}
 \eeqt
where $\Ra_T$ and $\Ra_m$ are the thermal and mass Rayleigh number, respectively, $\Pran_T$ the Prandtl number, and $\Lew_T$ the Lewis number (table~\ref{tab1:dimensionlessnumber}).

The stability of the system that exhibits a diffusive and a finger regime depends on both Rayleigh number types. In the finger regime, a small perturbation at the interface between the layers develops into a pattern of descending fingers, for instance salt fingers.

Another buoyancy-driven convection is the Rayleigh-B\'enard convection that arises by fluid in a reservoir heated from below. Convection results from thermal gradient. The set of equations is as follows:
\beqt
\label{eqRBC}
\begin{aligned}
\frac{\partial \mathbf{u}}{\partial t} {+} (\mathbf{u}\bcdot\bnabla)\mathbf{u} {-} \Pran_T\,\nabla^2 \mathbf{u}
  {+} \Pran_T\,\bnabla p &= {-} \Pran_T\,\Ra_T\,T\,{\mathbf j}, \\
\bnabla\bcdot\mathbf{u} &= 0, \\
\frac{\partial T}{\partial t} + \mathbf{u}\bcdot\bnabla T  - 
  \nabla^2 T            &=0 , \\
T=1 \textrm{ at bottom}, \quad T=0 \textrm{ at top}.
 \end{aligned} 
 \eeqt
$\Ra_T$ and $\Pran_T$ are the Rayleigh and Prandtl number, respectively. The balance between the gravitational and viscous forces is expressed by the Rayleigh number $\Ra_T$. When $\Ra_T$ is larger than a critical value that can be obtained analytically, convective patterns appear \cite{bodenschatz_recent_2000}.

\begin{figure}
\centering
\subfigure[]{\includegraphics[width=\linewidth]{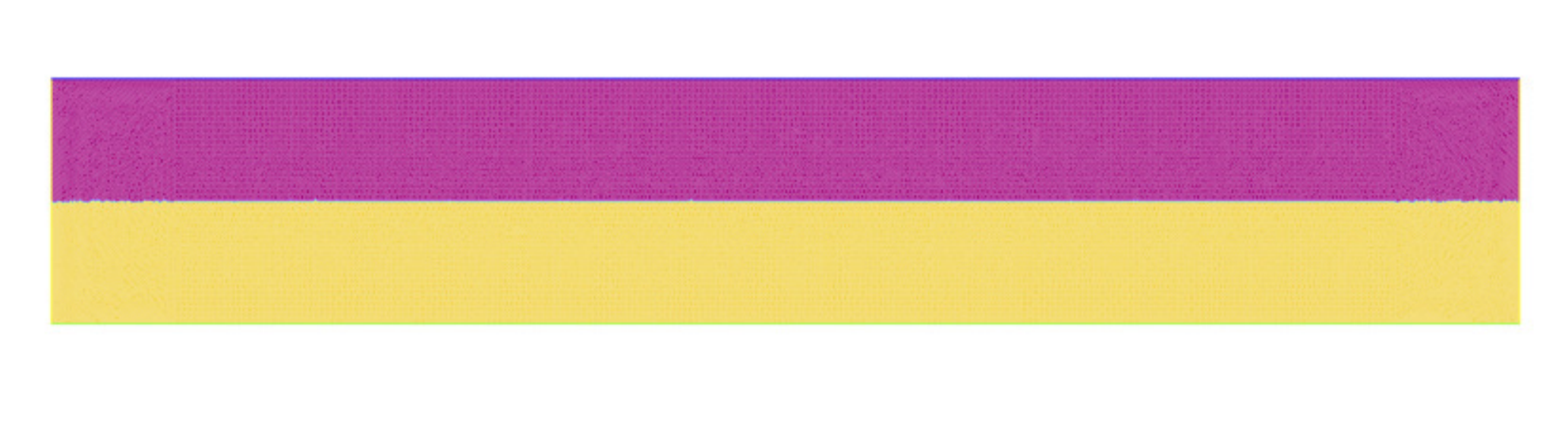}}
      \subfigure[]{\includegraphics[width=\linewidth]{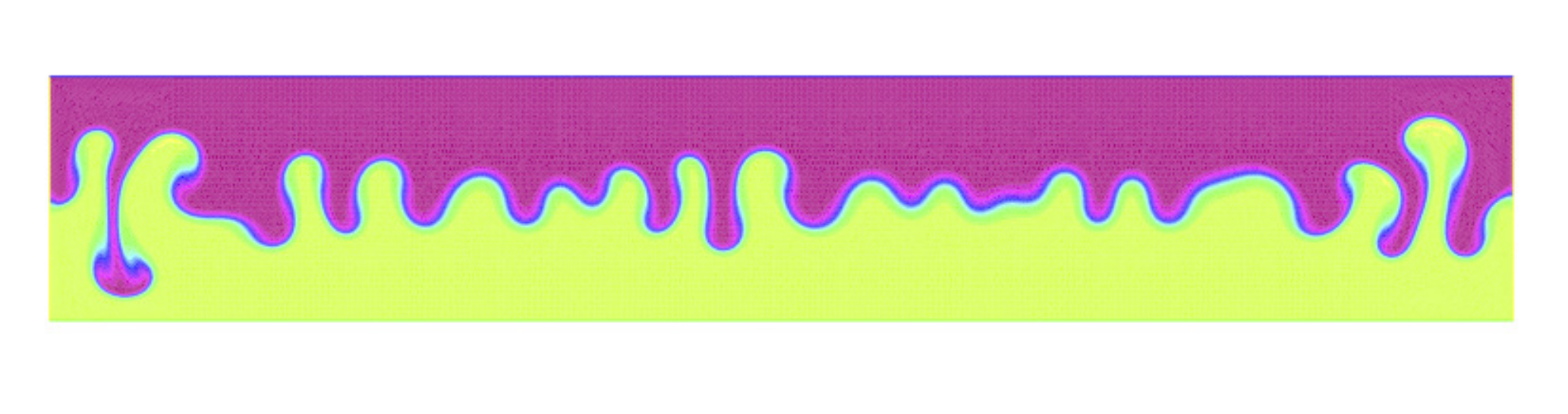}}
\caption{Plot of the solution $s$ in the double diffusive system (\ref{eqDDC}). Red area designates the region rich in $s$, while yellow area is poor in $s$. Descending plumes rich in $s$ and ascending plumes poor in $s$ form at the interface of two layers of fluid. The green color designates a region where the concentration of $s$ is slightly higher than that in the yellow area due to diffusion. (a) Initial condition: $s(x,y,t=0)={2}/{7}$, if $y<0.5$, and $s(x,y)=1.0$, if $y\geq0.5$. (b) $s$ at time $t$=0.1 for $\Ra_T=8000$, $\Ra_m={\Ra_T}/{6.2}$, $\Pran_T=7$ and $\Lew_T=0.01$. }
\label{fig:fig:DDC}
\end{figure}

\begin{figure}
\centering
\includegraphics[width=\linewidth]{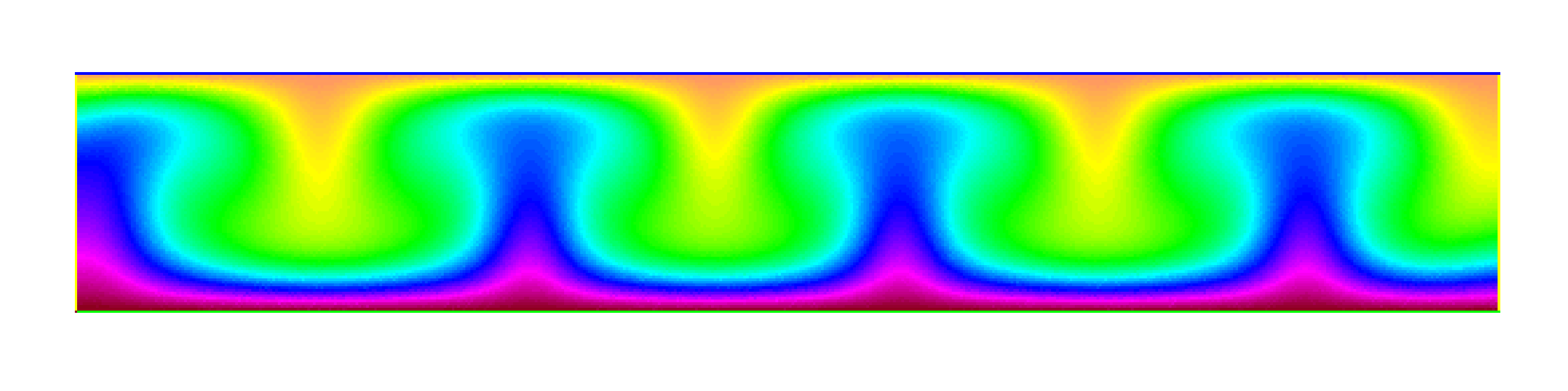}
\caption{Plot of the solution $T$ in the Rayleigh-B\'enard system (\ref{eqRBC}). Descending plumes of lower temperature and ascending plumes of higher temperature form in the fluid. $\Ra_T=6000$, $\Pran_T=50$.}
\label{fig:fig:RBC}
\end{figure}

The double diffusive convection equations in (\ref{eqDDC}) and the Rayleigh-B\'enard convection equations in(\ref{eqRBC}) are solved using our numerical method (figures~\ref{fig:fig:DDC} and \ref{fig:fig:RBC}). Plumes formed in the Rayleigh-B\'enard convection process are similar in shape to plumes of the chemotaxis--diffusion--convection. However, double diffusive and Rayleigh-B\'enard convections exhibit both ascending and descending plumes, whereas chemotaxis--diffusion--convection is only characterized by descending plumes.

%Fig16--------------------------------------------------------------------------------
%-------------------------------------------------------------------------------------

%Fig17--------------------------------------------------------------------------------
%-------------------------------------------------------------------------------------

In each problem, the effective Rayleigh numbers depend on the temperature difference between the opposite fluid domain surfaces, the gradients of temperature and salinity between the fluid layers, and the difference of density between the bacteria and solvent, for the Rayleigh-B\'enard, double diffusive, and chemotaxis--diffusion--convection systems, respectively. All of the three diffusion--convection processes have similar and distinct features (table \ref{tab8:convsystem}). The involved dimensionless parameters are listed in table \ref{tab1:dimensionlessnumber}.

%Table8--------------------------------------------------------------------------------
\begin{table*}
\renewcommand{\arraystretch}{1.5}
\caption{Recapitulative of the physical mechanisms involved in the double diffusive convection (DDC), chemotaxis--diffusion--convection (CDC) and the Rayleigh-B\'enard convection (RBC).}
\label{tab8:convsystem}
\centering
\resizebox{\linewidth}{!} {
    \begin{tabular}{c|c|c|c}
    
    ~                                  & DDC                                                       & CDC                                                             & RBC                                                    \\  \hline
    Hydrodynamics
    	& $\bnabla \bcdot u = 0$
		& $\bnabla \bcdot u = 0$
			& $\bnabla \bcdot u = 0$\\
			
    buoyancy $ (\rho V g)$
    	& $\begin{array}{c}  \dfrac{\mathrm{D}\mathbf{u}}{\mathrm{D}t}{=}  \Pran_T \nabla^2 \mathbf{u} {-} \Pran_T \bnabla p \\ -  \Pran_T \, ( \Ra_m \, s {-} \Ra_T T) {\mathbf j} \end{array}$ 
    		& $\begin{array}{c}\dfrac{\mathrm{D}\mathbf{u}}{\mathrm{D}t}{=} \Pran_{\tau} \nabla^2 \mathbf{u} - \Pran_{\tau}  \bnabla p \\  - \Ra_{\tau} \, \Pran_{\tau} \, n \, \, {\mathbf j}\end{array}$                  
    			& $\begin{array}{c}\dfrac{\mathrm{D}\mathbf{u}}{\mathrm{D}t}{=}\Pran_T \, \nabla^2 \mathbf{u} {-}  \Pran_T \, \bnabla p  \\ -  \Pran_T \, \Ra_T \, {\mathbf j}\end{array}$ \\   \hline

    Diffusion                              
    	& $ \arraycolsep=1.4pt\def\arraystretch{2.5} \begin{array}{rcl}\dfrac{\mathrm{D}T}{\mathrm{D}t}&=& \nabla^2 T \\ \dfrac{\mathrm{D}s}{\mathrm{D}t}&=&  \Lew_T \nabla^2 s \end{array}$ 
		& $\dfrac{\mathrm{D}n}{\mathrm{D}t}=  \nabla^2 n$ 
			& $\dfrac{\mathrm{D}T}{\mathrm{D}t}=  \nabla^2 T$  \\    \hline

    Chemotaxis                             
    	& $\O$                                                                                 
		& $\arraycolsep=1.4pt\def\arraystretch{2.5} \begin{array}{rl}\dfrac{\mathrm{D}n}{\mathrm{D}t}&=  - \mathsf{S} \bnabla \bcdot (n \, \bnabla c) \\ \dfrac{\mathrm{D}c}{\mathrm{D}t}&= \Lew_{\tau} \nabla^2 c  - \mathsf{H} n \end{array}$ 
			& $\O$\\    \hline

    Convection                             
    	&  $\arraycolsep=1.4pt\def\arraystretch{2.5} \begin{array}{rl}\dfrac{\mathrm{D}T}{\mathrm{D}t}&= \mathbf{u} \bcdot  \bnabla T \\ \dfrac{\mathrm{D}s}{\mathrm{D}t}&= \mathbf{u} \bcdot  \bnabla s \end{array}$ 
		& $\dfrac{\mathrm{D}n}{\mathrm{D}t}=  \mathbf{u} \bcdot \bnabla n$ 
			& $\dfrac{\mathrm{D}n}{\mathrm{D}t}=  \mathbf{u} \bcdot \bnabla n$\\    \hline

    I.C. for fingers                                   
    	& Layers                                                                                 
		& Any                                                                                           
			& Layers\\    \hline

    B.C. for mass                          
    	& Neumann                                                                           
		& Neumann                                                                                    
			& Dirichlet\\    \hline

    Physical regime                                 
    	& Diffusive \&                                                               
		& Diffusive,                                                           
			& Diffusive \&
			\\  
    ~                                 
    	& convective                                                                  
		& convective \&                                                     
			& convective\\  
    ~                                 
    	&   ~                                                        
		& chemotactic                                                           
			& ~ \\
	
    \end{tabular}
    }
\end{table*}
%-------------------------------------------------------------------------------------

Figure \ref{fig:convcells} shows that the arrangement of the convection cell structure can be the same for the three systems mentioned above. A particular position of the domain will be in a clockwise cell in some simulations and counter-clockwise cell in others. This behavior is also observed in Rayleigh-B\'enard convection simulations.

In the chemotaxis--diffusion--convection system, chemotaxis plays an essential role in the early stage, as it organizes the fluid domain. In a rectangular domain, starting from a given initial condition, aerotaxis of bacteria builds quasi-homogeneous layers in the horizontal direction, similarly to the double diffusive \cite{turner_double-diffusive_1974}. The multi-layered fluid creates a density gradient between the top of the stack layer and the bottom of the depletion layer that is
similar to the temperature gradient set by the Dirichlet boundary condition imposed in the Rayleigh-B\'enard case. The chemotaxis--diffusion--convection system evolves itself to a proper condition that leads to instabilities. 

Chemotaxis, that is not present in the other systems, brings flexibility in the choice of initial as well as boundary conditions. An analogy between the differential systems should stimulate further mathematical analysis of the system and a better understanding of the role of chemotaxis in convection problems. 

\section{Concluding remarks}
\label{conclusion}

In this paper, we have studied the chemotaxis--diffusion--convection system with the focus on the differential system rather than on the experimental settings. Our studies are based on the numerical implementation of the upwind finite element method with inconsistent Petrov-Galerkin weighted residual scheme to solve the coupled convective chemotaxis-fluid equations. Several simulation examples have been presented. They exhibit physically different spatially organized convection patterns.

The chemotaxis--diffusion--convection system can be described by three regimes. In the convective regime, the formation of plume patterns proceeds a temporal sequence of development stages. First, a spatial layered structure is created. Bacteria agglomerate in the upper stack layer, thus forming a depletion layer, where the bacterial density is very low. Subsequently, bacterial convection strengthens with time and instabilities in the stack layer appear. Finally, plumes of bacterial falling in the fluid are generated. The growth rate of the plumes is of the exponential type. The wavelength of the growing instabilities is defined from the parameters of the problem. Initial conditions appear to have a small influence on the plume number. The location of plumes can only be predicted when a very simple initial condition is set up. 

In the diffusive and chemotactic regimes, the chemotaxis system has a stabilizing effect on the fluid. When the taxis Rayleigh number increases, the gravitational force becomes dominant and instabilities occur and grow. From the phenomenological analysis and the numerical results, the critical Rayleigh number increases proportionally with the product of the chemotaxis sensitivity and head, when the chemotaxis sensitivity is large. 

In all buoyancy-driven flows (chemotaxis--diffusion, double diffusive, and Rayleigh-B\'enard convection), the dimensionless differential systems and the convection patterns are similar. Analogy between these types of convection should launch further analytical studies of the chemotaxis--diffusion--convection problem.

%%% Acknowledgments %%%%%%%%
\section*{Acknowledgments}
Y.D. received a PhD grand from Pierre and Marie Curie University - Sorbonne University and was also supported by the CASTS at National Taiwan University.




  \bibliographystyle{elsarticle-num} 
  \bibliography{cdconvection}


%\begin{thebibliography}{00}
%%% Text of bibliographic item
%
%\bibitem{}
%
%\end{thebibliography}
\end{document}